\documentclass{aa}
\usepackage{graphicx}
\def\ssim{\setbox0=\hbox{$\propto$}%
\setbox1=\hbox{$<$}\dimen0=\ht1%
\advance\dimen0by-1.2pt\,\lower.6\dimen0%
\copy0\kern-\wd0\raise.4\dimen0\copy1 \,}

\def\gsim{\setbox0=\hbox{$\propto$}%
\setbox1=\hbox{$>$}\dimen0=\ht1%
\advance\dimen0by-1.2pt\,\lower.6\dimen0%
\copy0\kern-\wd0\raise.4\dimen0\copy1\,}

\def\lambdab{\lambda\mkern-9mu\lower1.2pt\hbox{$\mathchar'26$}}%

\begin{document}
   \title{Stellar evolution with rotation XI: }

\subtitle{Wolf-Rayet star populations at different metallicities}

 \author{G. Meynet, A. Maeder}

     \institute{Geneva Observatory CH--1290 Sauverny, Switzerland\\
              email:  Georges.Meynet@obs.unige.ch\\
              email: Andre.Maeder@obs.unige.ch }

   \date{Received  / Accepted }

\abstract{Grids of  models of massive stars ($M \ge$ 20 $M_\odot$) with rotation are computed 
for metallicities $Z$ ranging from that of the Small Magellanic Cloud (SMC) to that of the Galactic Centre. The hydrostatic 
effects of rotation, the 
rotational mixing and the enhancements of the mass loss rates by rotation are included.
The evolution of the surface rotational velocities of the most massive O--stars 
mainly depends on the mass loss rates and thus on the initial $Z$ value.
The minimum initial mass for a star for entering the Wolf--Rayet (WR) phase is lowered by rotation. 
For  all  metallicities, rotating stars enter the WR phase at an earlier stage of  evolution
and the WR lifetimes are increased, mainly as a result of the
increased duration of the eWNL phase.
Models of WR stars predict in general rather low rotation velocities 
($ < 50$ km s$^{-1}$) with a few possible exceptions, particularly at metallicities lower
than solar where WR star models have in general faster rotation and  more chance to reach the
break--up limit.
The properties of the WR populations as predicted by the rotating models
are in general in much better  agreement with the observations in nearby galaxies.
Some possible remaining difficulties in these comparisons are mentioned. 
The evolution of the chemical abundances is largely influenced by rotation in all 
phases from the MS phase to the WN and WC phases. We also show that the interval of
initial masses going through the LBV stage is changing with $Z$ and $\Omega$.

The observed variation  with metallicity of the fractions of type Ib/Ic supernovae with respect to 
type II supernovae  as found by Prantzos \& Boissier 
(\cite{Pr03}) is very well 
reproduced by the rotating models, while  non--rotating models predict much too low ratios. 
This indicates that the minimum initial masses of single stars going through a WR phase
 are consistently predicted.
At $Z$ = 0.040, stars with initial masses above  50 $M_\odot$ reach a final mass at 
the time of supernova explosion between 5 and 7.5 $M_\odot$, while at $Z$ = 0.004,
like in the SMC, the final masses of stars are 
in the range of  17 -- 29 $M_\odot$. On the whole, rotation appears to be an essential parameter
even for the WR properties. Detailed tables describing the evolutionary tracks are available on the web. 
 }

   \maketitle
%

\section{Introduction}

Wolf--Rayet stars are considered to be bare stellar cores whose original H--rich envelopes 
have been removed either by strong stellar winds or by mass transfer through Roche Lobe 
Overflow in close binary systems (Conti \cite{Co76}; Chiosi \& Maeder \cite{Ch86} 
; Abbott \& Conti \cite{Ab87}). 
Their associations with young star forming regions implies 
that their progenitors must be massive stars (see e.g. 
the recent review by Massey \cite{Mas03} and references therein). WR stars 
have a deep impact on their surroundings thanks
to their high luminosity and their strong stellar winds. 
Their broad emission lines can be detected in the integrated spectrum of remote galaxies
(Kunth \& Sargent \cite{Ku81}; Schaerer et al. \cite{Sch99}) enabling us to study star formation and evolution in very different environments, 
from metal poor blue compact dwarf galaxies
to the vicinity of AGN (see e.g. L\'\i pari et al. \cite{Li03}). 
They contribute to the enrichment of the interstellar medium by newly synthesized elements. In particular their
winds at high metallicity may be heavily loaded with carbon (Maeder \cite{Maeder92}). Their winds may also eject significant amounts
of $^{26}$Al (see e.g. Vuissoz et al. \cite{Vu04}), responsible for the diffuse emission at 1.8 MeV  observed in the plane 
of our Galaxy (Prantzos \& Diehl \cite{Pr96}), in $^{19}$F (Meynet \& Arnould \cite{F19})
whose origin still remains largely unknown (Cunha et al. \cite{Cu03}), and in s--process elements (see e.g Arnould et al. \cite{Ar97}). 
The WC star winds are also
rich in $^{22}$Ne, which explains the high $^{22}$Ne/$^{20}$Ne isotopic ratio observed in the 
galactic cosmic ray source material (see e.g. Meynet et al. \cite{Mey01}).
WR stars are also the progenitors of type Ib/Ic supernovae (see the review by Hamuy \cite{Ham03}). Recently 
the spectrum of such a supernova was observed in the optical
transient of a $\gamma$--ray burst (see e.g. Hjorth et al. \cite{Hj03}), confirming the suspected link between these stars 
and the long $\gamma$--ray bursts (Woosley \cite{Wo93}).
For all these reasons Wolf--Rayet stars appear as objects worthwhile to be well understood.

In a previous paper (Meynet \& Maeder \cite{MMX}, paper X), we discussed the consequences 
of rotation on the properties of WR stars at solar metallicity. 
One of the main conclusions is that the theoretical predictions for the number ratios of Wolf--Rayet to
O--type stars, for the ratio of  WN to WC stars  and for the fraction of WR stars in the transition WN/WC phase, 
are in good agreement with the observations when the effects of rotation are
accounted for in stellar models.
In contrast, the models with present--day mass loss rates 
and no rotation do not succeed in reproducing the observed values. 
The main purpose of the present paper is to explore the case of metallicities lower and higher than 
solar and to see if the above conclusions still hold. 

Let us recall that interesting questions arise from the observations of WR stars both at low and high metallicity.
At low metallicity, it has generally been thought that WR stars might preferentially be formed by mass transfer
through Roche Lobe Overflow in close binary systems. For instance, it was thought that the majority, if not all the WR
stars in the SMC should be born thanks to the mass transfer mechanism.
The main reason is that, at low metallicity, the mass loss rates are much lower than at higher metallicity, 
thus making the ejection of the H--rich envelope by stellar winds
more difficult.
However, this idea was recently challenged by the works of Foellmi et al. (\cite{FoS03}, \cite{FoL03}). They
looked for periodic radial velocity variability in all the WR stars in the Small Magellanic Cloud and in two thirds of the WR stars in
the Large Magellanic Cloud. They found that the percentage of binaries among the WR stars is of the order of 40\%
for the SMC and 30\% for the LMC, thus comparable or even below the percentage of binaries among the WR stars in our Galaxy. 
This means that, in the SMC, at most 40\% of the WR stars could originate from
mass transfer through Roche Lobe Overflow (RLOF) in a close binary system. The real fraction is likely lower since 
RLOF has not necessarily occurred in all these systems.
Therefore even in the SMC, a large fraction of the WR stars likely originate via  the single star scenario. 

How is it possible ?
Does this mean that the mass loss rates are larger than usually found or 
that another process is at work which favours the evolution of massive stars into the WR phase ? 
In an earlier work (Maeder \& Meynet \cite{MM94}) we explored the first hypothesis, namely the effects of enhanced mass loss rates. 
We multiplied by a factor of two the mass loss rates given by de Jager et al. (\cite{Ja88}) during the O--type star phase and 
we adopted an average mass loss rate of 8$\cdot$10$^{-5}$ $M_\odot$ yr$^{-1}$ during the WNL phase, {\it i.e.} twice
the average mass loss rate given by Abbott \& Conti (\cite{Ab87}). Doing so, we obtained a good agreement 
with the observed WR populations for the metallicities between that of the SMC and up to twice the solar metallicity. 
However, nowadays clear evidence of clumping in stellar winds of massive stars (Nugis et al. \cite{Nu98}) has been found and the new estimates
of the mass loss rates (Nugis \& Lamers \cite{NuLa00}) 
are reduced by a factor 2 to 3 with respect
to the enhanced mass loss rates described above. Therefore the enhanced mass loss rate hypothesis is ruled
out and another process must be at work. 
In paper X, we showed that rotation might well be this process. Indeed,
in massive star models, rotation favours the evolution into the WR phase in two ways, 
first by allowing the star to enter into the WR phase at an earlier stage, thus making the WR lifetime longer and 
secondly by allowing smaller initial mass stars to go through a WR phase. These effects, as we shall see below, are also present at low
metallicity and allow us to reproduce the observed WR populations in the LMC and SMC without assuming that a large fraction
of WR stars owe their existence to mass transfer in close binary systems.

We shall also study the
effects of rotation at higher metallicity than solar, more precisely at twice the solar 
metallicity, a value often quoted for the galactic centre (although some authors quote for this region
a value of the metallicity similar to that of the solar neighbourhood, see Carr et al. \cite{Ca99}; Najarro \cite{Na03}). 
This region of the Milky Way is very rich in massive stars (according to Figer et al.~\cite{Fi02}, the Arches cluster
contains about 5\% of all known WR stars in the Galaxy) and it is thus
interesting to derive the properties of the WR stars predicted by the present rotating models.
Moreover, populations of Wolf--Rayet stars at a higher metallicity may also be of interest for studying
the stellar population in the vicinity of AGNs (see e.g. the review by Heckman \cite{Hec99}). 

Sect. 2 briefly summarizes the physics of the models. Evolution of the surface rotational velocities
is discussed in Sect. 3.  The evolutionary tracks are presented in Sect. 4.
The effects of rotation on WR star formation and WR lifetimes
at different metallicities
are discussed in Sect. 5. Comparisons with the observed WR populations are
performed in Sect. 6. Finally the predicted surface abundances of the present WR stellar models are discussed in Sect. 7.

\begin{figure*}[t]
  \resizebox{\hsize}{!}{\includegraphics[angle=-90]{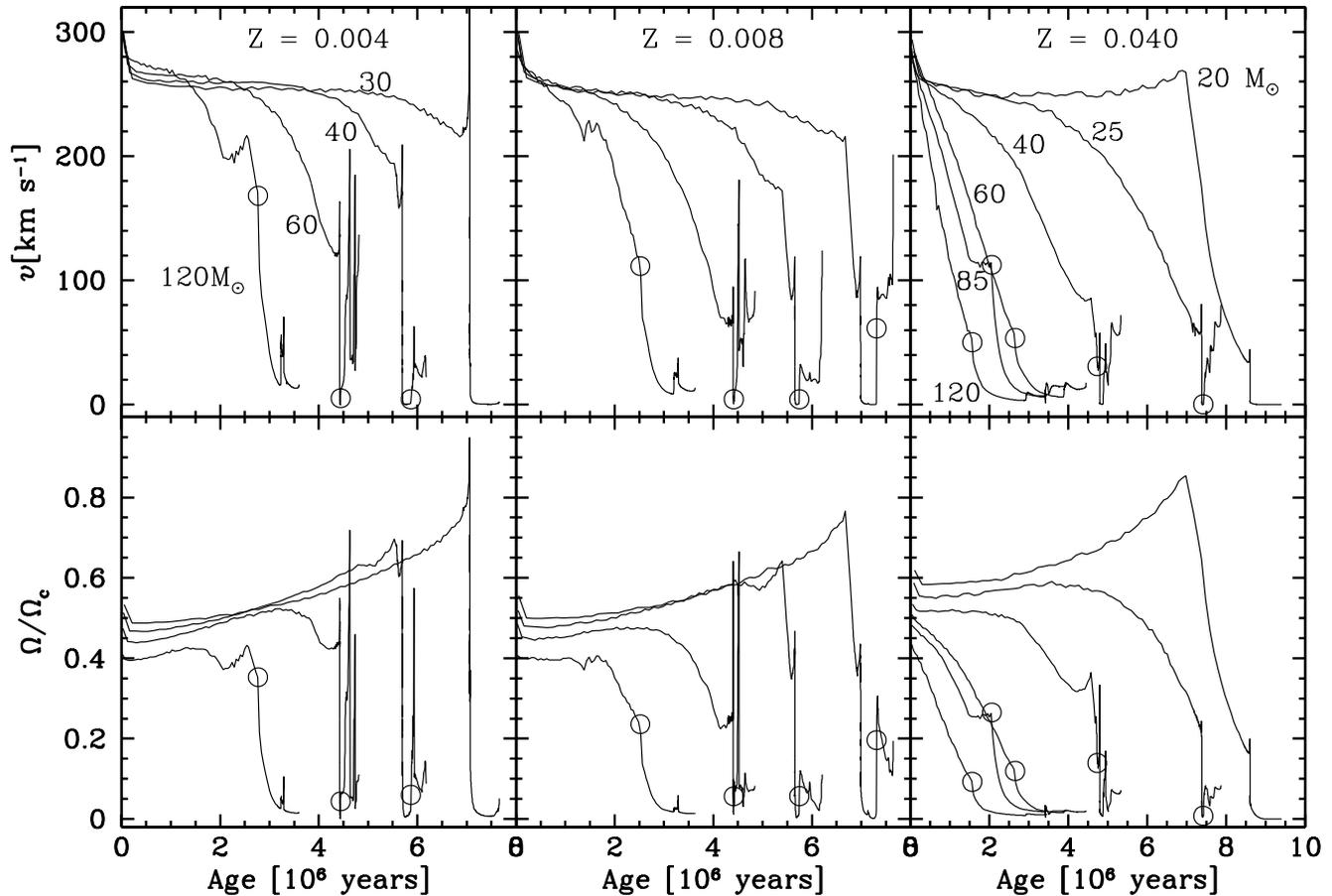}}
  \caption{Evolution of the equatorial velocities ({\it upper panels}) and of
  the ratio  $\frac{\Omega}{\Omega_{\mathrm{c}}}$ of the angular velocity to 
  the critical angular velocity ({\it lower panels}) at the surface of star models 
  of different initial masses and metallicities.
  Values of the initial masses are indicated
  in the upper left and right panels. The initial masses corresponding to the tracks plotted 
  in the middle panels
  are the same as those shown in the left panels. The initial rotational velocity of all the models
  is 300 km s$^{-1}$. Empty circles are placed at the stage when the star enters into the Wolf--Rayet phase.}
  \label{vrotZ}
\end{figure*}

\section{Physics of the models}

Most of the physical ingredients of the present models are the same as
in the solar metallicity models of Meynet \& Maeder (\cite{MMX}, paper X). They differ in only two points:
 
\begin{itemize}
\item  
The initial compositions are adapted for the different
metallicities considered here.
For a given metallicity $Z$ (in mass fraction), the initial helium mass fraction
$Y$ is given by the relation $Y= Y_p + \Delta Y/\Delta Z \cdot Z$, 
where $Y_p$ is the primordial
helium abundance and $\Delta Y/\Delta Z$ the slope of 
the helium--to--metal enrichment law. We use the same values as in Maeder \& Meynet (\cite{MMVII})
{\it i.e.} $Y_p$ = 0.23 and $\Delta Y/\Delta Z$ = 2.5.
For the metallicities $Z$ = 0.004, 0.008 and 0.040 considered in this work, we thus have
$X$ = 0.757, 0.744, 0.640 and $Y$ = 0.239, 0.248, 0.320 respectively.
For the heavy elements
we adopt the same mixture as the one
used to compute the opacity tables for solar composition. 
In that respect the present grid differs also from 
our previous
non--solar metallicity rotating models (papers VII and VIII). In these models, we adopted the  
``enhanced alpha elements'' opacity tables of 
Iglesias \& Rogers (\cite{Iglesias}) instead of the ``solar composition'' ones.
As the relative abundances of $\alpha$--nuclei vary according to the
galactic history, the present choice may be better.

\item Secondly the wind anisotropies induced by rotation were
neglected. This last choice appears justified in view of the results obtained in paper X. Indeed for the
initial velocities considered ($\upsilon_{\rm ini} = 300$ km s$^{-1}$), the effects of the wind
anisotropies have been shown to be very small.
Let us however emphasize that this is not true for higher initial velocities (Maeder \cite{Ma02}).
\end{itemize}

Since mass loss rates are a key ingredient of the models in the mass
range considered here, let us recall the prescriptions used.
The changes of the mass loss rates $\dot{M}$ with  
rotation are taken into account as explained in Maeder \& Meynet (\cite{MMVI}).
As reference mass loss rates 
we adopt the mass loss rates of Vink et al. (\cite{Vink00}; \cite{Vink01})
who take account of the occurrence of bi--stability
limits which change the wind properties and mass loss rates.
For the domain not covered by these authors
we use the empirical law devised by 
de Jager et al. (\cite{Ja88}).
Note that this empirical law, which presents
a discontinuity in the mass flux near the Humphreys--Davidson limit,
implicitly accounts for the mass loss rates of LBV stars.
For the non--rotating
models, since the empirical values
for the mass loss rates are based on 
stars covering the whole range of rotational velocities, 
we must apply a reduction factor to the empirical rates to make
them correspond to the non--rotating case. The same reduction factor was used as
in Maeder \& Meynet (\cite{MMVII}).
During the Wolf--Rayet phase we use
the mass loss rates by Nugis \& Lamers (\cite{NuLa00}). These mass loss rates,
which account for the clumping effects in the winds,  
are smaller by a factor 2--3 than the mass loss rates used in our previous 
non--rotating ``enhanced mass loss rate'' stellar grids
(Meynet et al. \cite{Mey94}). 

During the non--WR phases of the present models, we assumed
that the mass loss rates depend on the initial metallicity
as $\dot M(Z)=(Z/Z_\odot)^{1/2} \dot M(Z_\odot)$
(Kudritzki \& Puls \cite{KP00}; Vink et al. \cite{Vink01}).
For models at $Z=0.040$, 
we also compute a series of models
with metallicity dependent mass loss rates during the WR phase. According to Crowther et al.~(\cite{Cro02})
mass loss rates during the WR phase may show the same metallicity dependence as the
winds of O--type stars, {\it i.e.} scale with $\sim (Z/Z_\odot)^{1/2}$.

The effective temperature of Wolf--Rayet stars is a delicate problem, since the
winds may have a non--negligible optical thickness. Here we adopt 
a simple correction scheme to take account of this effect (e.g. Langer~\cite{La89}): the effective radius
$R_{\rm eff}$ at the optical thickness $\tau=2/3$ is related to the classical photospheric radius $R$
by the relation

$$R_{\rm eff}  = R + {3 K |\dot M| \over 8 \pi \upsilon_{\infty}},$$
where $K$ is the opacity and the other symbols have their usual meaning.
More details on the procedure to estimate $K$, $\tau$ and $\upsilon_{\infty}$
are given in Schaller et al. (\cite{Sch92}).
The effective temperature at $\tau=2/3$ is then obtained by the usual relation
$L=4\pi R^2_{\rm eff} \sigma T^4_{\rm eff}$. Such a correction
has been applied in the WR stages and only there.

A moderate overshooting is included in the present
rotating and non--rotating models. The radius of the convective cores are increased
with respect to their values obtained by the Schwarzschild criterion by a quantity
equal to 0.1 H$_{\rm p}$, where H$_{\rm p}$ is the pressure scale height estimated at the Schwarzschild
boundary. 
The effect of rotation on the transport of the chemical species and of the
angular momentum are included as in our papers VII and VIII.

As initial rotation, we have considered a value equal to 300 km s$^{-1}$
on the ZAMS for all the initial masses and metallicities considered. At solar metallicity, this
initial value produces time--averaged equatorial velocities on the MS well in the observed range,
i.e. between 200 and 250 km s$^{-1}$. At low metallicities this initial rotational velocity corresponds also to mean values
between 200 and 250 km s$^{-1}$ on the MS phase, while at twice the solar metallicity, the mean velocity is lower, between
160 and 230 km s$^{-1}$ (see Table~\ref{tbl-1}). Presently we do not know the distribution of the
rotational velocities
at these non--solar metallicities and thus we do not know if the adopted initial velocity corresponds to the average observed values. 
It may be that at lower metallicities the initial velocity distribution contains a larger number of
high initial velocities (Maeder et al.~\cite{MG99}), in which case 
the effects of rotation described below would be underestimated at low metallicity.

All the models were computed 
up to the end of the helium--burning phase. Their further evolution in the advanced stages
will be presented in a forthcoming paper (Hirschi et al. in press).
In order to facilitate future detailed comparisons, we provide electronic tables describing the evolutionary tracks presented
in this paper\footnote{The tables can be found at the web address: {\it http://www.unige.ch/sciences/astro/an/}; choose on the
left ``RESEARCH GROUPS'', then under the title ``EVOLUTION STELLAIRE'', choose ``ETOILES MASSIVES'', 
then ``Tables des resultats des recents modeles avec rotation''.}. 
For each initial mass, we extracted
350 points describing the whole sequence from the ZAMS to the end of the core He--burning phase.
At each time step we provide the following quantities: age in years,
actual mass in solar masses, log $L/L_\odot$, log $T_{\rm eff}$
(which takes account of the optical thickness of the winds for WR stars, see above), 
the surface abundances in mass fraction of H,
He, $^{12}$C, $^{13}$C, $^{14}$N, $^{16}$O, $^{17}$O, $^{18}$O, $^{20}$Ne, $^{22}$Ne,
the fraction of the mass of the star occupied by the convective core,
the uncorrected effective temperature, {\it i.e.} which does not take account of the
optical thickness of the winds (for WR stars only),
the mass loss rate in solar masses per year,
log $\rho_{\rm c}$, the central density, log $T_{\rm c}$, the central temperature,
the abundances at the centre of the star in mass fraction of the same elements as above,
the ratio of the polar radius to the equatorial radius, the surface equatorial velocity in km s$^{-1}$,
and the ratio of the surface angular velocity to the break--up velocity.
For the nuclear burning phases, models with the same number in the tables 
always correspond to the same evolutionary stage, {\it i.e.} have
the same mass fraction of hydrogen or helium at the centre: point 1 corresponds to the ZAMS stage,
point 100
to the end of the core H--burning phase, point 201 and 350 to the beginning and end respectively of
the core He--burning stage.

To conclude this section, let us emphasize that
the amplitudes of the effects studied in this paper are particularly sensitive to the mass loss rates, the
extension of the convective cores and the initial rotation.
Does this mean that the present results 
are very model--dependent~? We do not think so for the following reasons: firstly 
we are not free to change the above parameters
beyond certain limits, for instance the initial rotation was chosen in order to reproduce the averaged
observed velocity in the MS band, mass loss comes in part from the radiation wind theory, in part from
empirically based relations, the overshoot is constrained from comparisons between the observed and
computed MS width; secondly 
the present physical ingredients
already allowed us to reproduce many observed features such as the observed surface enrichments at different metallicities as well as the
the blue to red supergiant ratio in the SMC. These good agreements give some support to the chosen set of physical ingredients; 
thirdly we think that, in this first extensive study of the effects of rotation
on the WR formation at different metallicity,
it is necessary to focus on the effects of rotation alone, all other physical ingredients being kept the same.
In the future, improvements of the physical ingredients of the models will change some of the results but likely
the effects of rotation discussed here will qualitatively remain the same.
In that respect it is interesting to mention that in Meynet (\cite{Me00}) we deduced
from a previous grid of Wolf--Rayet stellar models at solar metallicity (paper V) the theoretical value
for the number ratio of WR to O--type stars. The results obtained
(WR/O = 0.026 for $\upsilon_{\rm ini}=0$ km s$^{-1}$ and 0.072 for  $\upsilon_{\rm ini}=300$ km s$^{-1}$)
are very similar to those obtained from the present grid (WR/O = 0.02 for $\upsilon_{\rm ini}=0$ km s$^{-1}$ 
and 0.07 for  $\upsilon_{\rm ini}=300$ km s$^{-1}$,
see Table~\ref{tblwr}), despite the fact that the two sets of stellar models differ in many respects.
Among other things, they differ by the prescription
for the mass loss rates, the amount of overshooting and the form of the shear diffusion coefficient.
In our view, the nearly identical results obtained in the two cases supports 
the idea that the results concerning the WR populations are not very sensitive
to small changes of these parameters.

\section{Evolution of the rotational velocities at various metallicities}

\begin{figure}[t]
  \resizebox{\hsize}{!}{\includegraphics{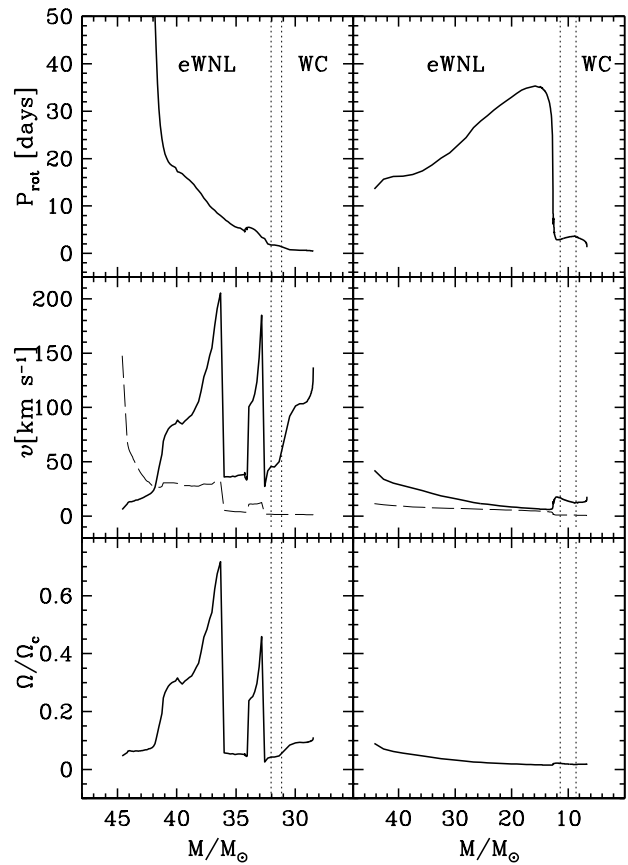}}
  \caption{Evolution as a function of the actual mass 
  of the rotation period, of the surface equatorial velocity
  and of the ratio of the angular velocity to the critical value during
  the WR stage of rotating stars. The long--dashed lines
in the panels for the velocities show the evolution of the radius in solar units.
{\it Left}: the WR phase of a star with an initial
  mass of 60 $M_{\odot}$ with $v_{\mathrm{ini}}= 300$ km~s$^{-1}$ and $Z$ = 0.004. 
{\it Right}: for an initial mass of 60 $M_{\odot}$ with $v_{\mathrm{ini}}= 300$ km~s$^{-1}$
and $Z$ = 0.040.}
  \label{vrotwr1}
\end{figure}

As discussed in paper X, the evolution of the rotational 
velocities at the stellar surface depends mainly 
on two factors, the internal coupling and the mass loss. The internal coupling is
achieved in radiative zones by meridional circulation and by shear diffusion.
During the Main Sequence phase, the internal coupling transports angular momentum
from the contracting core to the
expanding envelope and thus maintains the surface angular velocity above 
the value it would have in case of no--coupling {\it i.e.} of local conservation of the angular momentum. 
Mass loss on the other hand removes the angular momentum contained in
the ejected outer surface layers. Thus,
the two effects act in opposite directions for the evolution
of the surface velocities. The stronger the coupling mechanisms, the higher the values of the surface velocity
at a given stage. The stronger the mass loss rates,
the faster the decline of the surface velocities. Let us add that polar winds may decrease the quantity of angular momentum lost by stellar winds; 
however, as recalled in Sect.~2, we did not take account of this
effect due to  the moderate initial velocities considered.

How do these two processes, internal coupling and mass loss, depend on the initial metallicity ?
For what concerns mass loss, 
it is now well established both theoretically and observationally that the lower the metallicity,
the lower the mass loss rates (Kudritzki \& Puls \cite{KP00}; Vink et al. \cite{Vink01}). 
This metallicity effect alone would produce a slower decline of the surface
velocity at lower metallicity. Now what about the metallicity dependence of the internal coupling mechanisms ?
At lower metallicities, stars are more compact, and thus
the densities in their outer layers at a given stage are higher. 
Since
meridional circulation is more efficient 
than shear
for the transport of angular momentum (Meynet \& Maeder \cite{MMV}), and since 
in the outer layers the driving effect for the meridional velocity is the ``Gratton--\"Opik'' 
term which is proportional to the inverse of the density, the outward transport of angular momentum in these layers will be less efficient 
at lower than at higher metallicity. This clearly reduces the internal coupling, and
weakens the value of the surface velocity obtained at a given stage, all
other things being equal. Thus when metallicity decreases, on one side, less angular momentum is removed by stellar winds, and on the other side, less angular momentum is brought up to the surface by the internal transport processes.
Taken together these two effects tend to decrease the quantity of angular momentum
removed by stellar winds at low metallicity (see also Maeder \& Meynet \cite{MMVII}, paper VII).

For the evolution
of the surface velocities, things are a little more complicated:
at lower metallicity, the metallicity dependence of mass loss rates favours
higher values for the surface velocities at a given stage. On the other hand, the less efficient internal coupling favours lower values
for the surface velocities. 
The numerical models of the present work show that for the most massive stars
(M $>$ 30 $M_\odot$),
the effect of the metallicity dependence of the mass loss rates dominates.
This can be seen for instance in the case
of the 40 $M_\odot$ tracks plotted in Fig.~\ref{vrotZ}: at high metallicity ($Z=0.040$) the surface velocity 
rapidly declines during the core H--burning phase, while at low metallicity ($Z=0.004$) 
the star keeps a high surface velocity during most of this phase. 

The evolution of the ratio $\Omega/\Omega_c$ of
surface angular velocity to the break--up or critical angular velocity is also very different
at high and low metallicity. At a low metallicity, the ratio $\Omega/\Omega_{\rm c}$ at the surface of the 60 $M_\odot$ model
remains near the value of 0.5 during a great part of the Main Sequence phase, while at high metallicity the ratio
continuously decreases.

As was already emphasized in paper X,
we again stress that any comparison between observed and predicted rotation for the large masses
(M $>$ 30 $M_\odot$) is really much more a test bearing on  the mass loss rates than a test of
the internal coupling and evolution of rotation. 
For instance, the increase of $\Omega/\Omega_{\rm c}$ at low metallicity was more pronounced in our 
previous low metallicity grids (papers VII and VIII). This is because the mass loss rates adopted in these
previous grids were lower by about a factor of two than the new values adopted here.
This well illustrates both the uncertainties still pertaining to the mass loss rates and
the sensitivity of the evolution of the surface velocity  on this still not very well known physical ingredient of the stellar models.

\begin{figure*}[t]
  \resizebox{\hsize}{!}{\includegraphics[angle=-90]{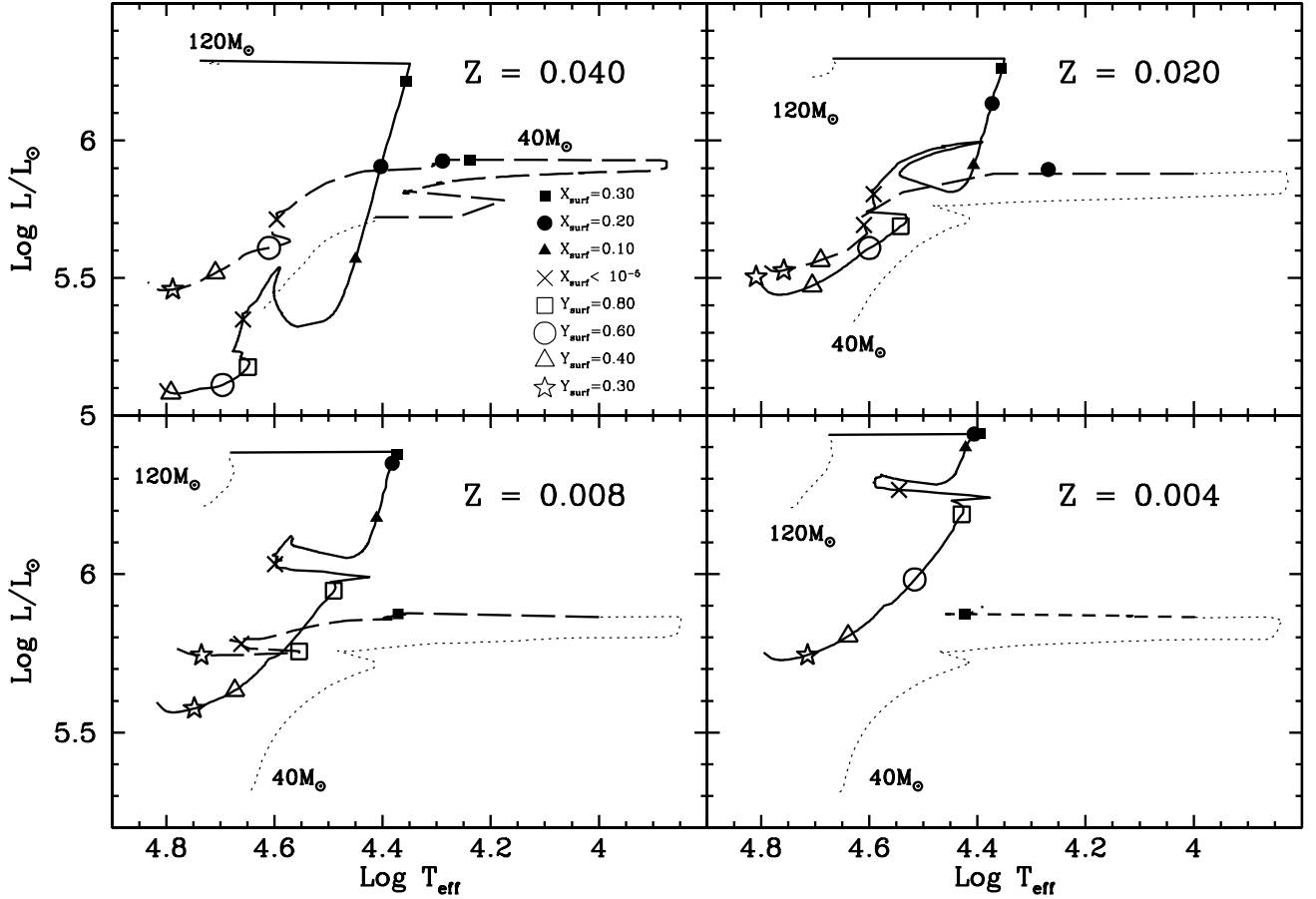}}
  \caption{Evolutionary tracks for 40 and 120 M$_\odot$ rotating  models at different metallicities.
  The initial velocity is 300 km s$^{-1}$. 
  The light dotted lines correspond to the non--WR part of the tracks. The tracks
during the WR phase are shown by heavy lines (continuous for the 120 M$_\odot$ and
dashed for the 40 M$_\odot$ model). Symbols along the tracks are placed where the indicated
surface hydrogen ($X_{\rm surf}$) and helium ($Y_{\rm surf}$) abundances are reached.
}
  \label{dhr004}
\end{figure*}

\subsection{Evolution of the rotational velocities during the WR stages}

From Fig.~\ref{vrotZ} one can see that after the entry into the WR phase, the surface velocity can
show rapid changes. These abrupt variations of the surface velocities are linked to rapid changes  
in the position of the star in the Hertzsprung--Russel diagram caused by surface 
composition modifications induced by both rotational diffusion and mass loss by stellar winds.

Fig.~\ref{vrotwr1} shows specifically
the evolution during the WR stages of the rotation period $P$ ($2\pi/\Omega$), of the
rotation velocities $v$ at the equator and of the fraction 
$\frac{\Omega}{\Omega_{\mathrm{c}}}$ of the angular velocity to 
the critical angular velocity at the surface 
of a star model with an initial mass of
60 $M_{\odot}$ and $v_{\mathrm{ini}}= 300$ km s$^{-1}$
at low and high metallicities ({\it left panels}, Z = 0.004; {\it right panels}, Z = 0.040).
One sees that at low metallicity, much greater and more rapid changes of the surface velocities
are expected than at higher metallicity. Thus the encounter of the break--up limit during the WR phase 
may appear more often at lower metallicity (see the
lower panels in Fig.~\ref{vrotwr1}).
This comes from the fact that at low metallicity the whole WR phase
occurs while the star is in the core He--burning phase, a phase during which
our most massive star models experience 
strong changes in radius.
In contrast, at $Z=0.040$, nearly the whole eWNL phase occurs 
during the core H--burning phase (for a definition of the eWNL phase see Sect.~5.1).
One sees that the high mass loss rates undergone by the star makes the surface velocity to decrease regularly.
Only when the star has lost its complete H--rich envelope
(i.e. when the star enters the eWNE
stage in terms of evolutionary models), does strong contraction of the surface layers,
caused by the decrease of the opacity in the outer regions,
produces a small increase
of the surface velocity.
We see that 
despite the strong increase of $\Omega$ the velocity $v$  and 
$\frac{\Omega}{\Omega_{\mathrm{c}}}$ do not change
very much, since the radius is much smaller in the eWNE stage.
All the transitions in the WR stage are
fast, thus the transfer of angular momentum
by meridional circulation is limited and the evolution of the rotation at the surface
is dominated by the local conservation of angular momentum, which explains 
the fast and large decreases of $P$ or $\Omega$. 

One notes that in general
the final surface velocity at low metallicity is also higher than at
high metallicity, a fact which may be easily explained as a consequence of the lower mass loss rates
at lower metallicity. From Table~\ref{tbl-1}, one sees also that
at a given metallicity the velocities obtained at the end of the He--burning phase
for the WR stars are in general smaller for the higher initial mass stars.
This results from the fact that the higher the initial mass, the larger the amount
of mass (and thus of angular momentum) removed by the stellar winds.

The implications of the present models for the rotation rates of pulsars at birth
and the possible progenitors of the collapsar models will be considered in a forthcoming paper.
Let us just mention here that most of the WR stellar models have enough specific angular momentum
in their core to be good candidates for collapsar models. On the other hand,
if no angular momentum is lost during the core--collapse supernova explosion, these models would predict
very fast rotational velocities for the neutron stars at birth. These results are in qualitative agreement
with those obtained by Heger \& Woosley (\cite{HW04}), and Woosley \& Heger (\cite{WH04}) in the case of no magnetic field. 

\section{Evolutionary tracks, lifetimes and final masses}

In Fig.~\ref{dhr004} some evolutionary tracks of rotating models are shown.
Note that during the WR phase the values of the effective temperature used to draw the tracks take account
of the non--negligible optical thickness of the winds (see Sect.~1).
At high metallicity and for the most massive stars, the entry into the WR phase occurs 
during the MS phase (see Sect.~5 below). Therefore the non--WR part of the track is quite short. For lower
initial mass stars, rotation makes the tracks overluminous and more extended towards lower effective temperatures
during the MS phase. This was also the case in previous works (see Heger \& Langer \cite{he00}; Meynet \& Maeder \cite{MMV}, \cite{MMX}).
Note however that for very fast rotation the star may be so efficiently mixed that it has a nearly
homogeneous evolution and would present a very blue track during the MS phase (Maeder~\cite{Maeder87}).
This is the case for the 500 km s$^{-1}$ 60 M$_\odot$ stellar model
at Z=0.004 in the present grid (see Table~\ref{tbl-1}).
Let us recall also that at low metallicity, for those stars which do not already enter the WR phase
during the MS phase, the evolution towards the red supergiant stage
is favoured when rotation is included in the stellar model (Maeder \& Meynet \cite{MMVII}).

Table~\ref{tbl-1} presents some properties of the models. Columns 1 and 2 give the initial mass and the initial velocity $v_{\rm ini}$ respectively.
The mean equatorial rotational velocity $\overline{v}$ during the MS phase is indicated in column 3.
This quantity is defined as in paper V. 
The H--burning lifetimes $t_H$, the lifetimes as an O--type star on the MS $t_{\rm O}$
(we assumed that O--type stars have an effective temperature higher than about 33 000 K),
the masses $M$, the equatorial velocities $v$, the helium surface abundance $Y_s$ and the 
surface ratios (in mass fraction) N/C and N/O at the end of the H--burning phase are given in columns 4 to 10.
Columns 11 to 17 present some characteristics of the stellar models at the end of the He--burning phase,
$t_{He}$ is the He--burning lifetime. More details on the models are given in the electronic tables (see Sect.~2).

\begin{table*}
\caption{Properties of the stellar models at the end of the H--burning phase
and at the end of the He--burning phase. The masses are in solar mass, 
the velocities in km s$^{-1}$, the lifetimes in million years and the abundances in mass fraction.} \label{tbl-1}
\begin{center}\scriptsize
\begin{tabular}{ccc|ccccccc|ccccccc}
\hline
    &         &         &         &         &     &      &      &      &     &      &      &      &         &   & &         \\

M & $v_{\rm ini}$ & $\overline{v}$ & \multicolumn{7}{|c|}{End of H--burning} &\multicolumn{7}{|c}{End of He--burning}      \\
    &         &         &         &         &     &      &      &      &     &      &      &      &         &   & &         \\
    &  &   & $t_H$ & $t_{\rm O}$ & M & $v$ & Y$_s$ & N/C & N/O &  $t_{He}$ & M & $v$ & Y$_s$ & C &  N & O \\
    &         &         &         &         &     &      &      &      &     &      &      &      &         &    & &        \\
\hline
    &         &         &\multicolumn{14}{ }{ }       \\
   &   &   & \multicolumn{14}{c}{\bf Z=0.004}\\
    &         &         &\multicolumn{14}{ }{ }       \\   
120 & 300 & 244 &  3.232 & 2.770 & 52.930 & 23.9 & 0.98 & 42.3 & 60.0 &  0.362 & 17.178 & 16.5  & 0.24 & 0.46 & 0 & 0.29     \\
60  & 300 & 236 &  4.422 & 4.206 & 51.824 & 157.0 & 0.41 & 5.33 & 1.82 &  0.383 & 28.465 & 136.8  & 0.14 & 0.34 & 0 & 0.51     \\
60$^1$  & 500 & 392 &  5.103 & 4.657 & 36.640 &  65.7 & 0.93 & 45.1 & 43.6 &  0.400 & 12.359 & 48.1  & 0.29 & 0.49 & 0 & 0.22     \\
40  & 300 & 249 &  5.690 & 5.114 & 36.755 & 209.3 & 0.29 & 1.64 & 0.53 &  0.476 & 22.333 & 22.0  & 0.74 & 4e-5 & 252e-5 & 14e-5     \\
30  & 300 & 250 &  7.059 & 6.196 & 28.654 & 336.4 & 0.25 & 0.99 & 0.31 &  0.590 & 18.866 & 2.59  & 0.63 & 9e-5 & 218e-5 & 47e-5     \\
    &         &         &\multicolumn{14}{ }{ }       \\
   &   &   & \multicolumn{14}{c}{\bf Z=0.008}\\
    &         &         &\multicolumn{14}{ }{ }       \\ 
120 & 300 & 222 &  3.199 & 2.522 & 36.612 & 12.4 & 0.96 & 46.5 & 52.2 &  0.427 & 13.393 & 13.8  & 0.25 & 0.47 & 0 & 0.27     \\
60  & 300 & 209 &  4.401 & 4.255 & 48.447 & 88.6 & 0.51 & 9.85 & 3.3 &  0.435 & 16.446 & 91.7  & 0.19 & 0.41 & 0 & 0.39     \\
40  & 300 & 233 &  5.651 & 4.928 & 35.294 & 11.6 & 0.34 & 2.59 & 0.8 &  0.550 & 17.342 & 123.9  & 0.26 & 0.30 & 6.9e-4 & 0.43     \\
30  & 300 & 240 &  6.982 & 5.817 & 27.457 & 119.2 & 0.28 & 1.39 & 0.4 &  0.658 & 12.106 & 201.2  & 0.59 & 0.21 & 2.9e-3 & 0.19     \\
    &         &         &\multicolumn{14}{ }{ }       \\   
   &   &   & \multicolumn{14}{c}{\bf Z=0.040}\\
    &         &         &\multicolumn{14}{ }{ }       \\ 
120 & 0   &  0  &  2.533 & 1.823 & 18.569 & 0   & 0.94 & 52.5 & 25.8 &  0.467 & 8.572 & 0    & 0.37 & 0.46 & 0 & 0.12
   \\
120 & 300 & 157 &  2.929 & 1.565 & 13.954 & 4.9 & 0.93 & 56.1 & 23.5 &  0.525 & 7.112 & 9.0  & 0.39 & 0.46 & 0 & 0.10     \\
120$^2$ & 300 & 157 &  3.182 & 1.565 & 9.595 & 4.0 & 0.92 & 60.1 & 22.4 &  0.673 & 4.846 & 7.9  & 0.62 & 0.30 & 0 & 0.03     \\

85  & 300 & 180 &  3.413 & 2.061 & 15.592 & 8.7 & 0.89 & 58.8 & 19.4 &  0.525 & 7.295 & 18.1  & 0.38 & 0.46 & 0 & 0.11     \\
85$^2$   & 300 & 180 &  3.560 & 2.061 & 11.633 & 7.2 & 0.88 & 61.8 & 18.6 &  0.746 & 4.7087 & 16.6  & 0.42 & 0.44 & 0 & 0.09     \\
60  & 0   &  0  &  3.122 & 2.213 & 29.272 & 0   & 0.73 & 73.4 & 10.8 &  0.410 & 11.329 & 0    & 0.30 & 0.47 & 0 & 0.18
   \\
60  & 300 & 176 &  3.894 & 2.652 & 12.684 & 8.9 & 0.93 & 57.3 & 20.1 &  0.549 & 6.686 & 17.8  & 0.43 & 0.44 & 0 & 0.08     \\
60$^2$   & 300 & 176 &  4.080 & 2.652 & 9.292 & 7.1 & 0.92 & 59.9 & 19.5 &  0.682 & 4.772 & 14.6  & 0.65 & 0.28 & 0 & 0.02     \\
40  & 300 & 187 &  4.793 & 4.073 & 29.948 & 52.7 & 0.57 & 10.7 & 2.3 &  0.526 & 11.418 & 71.9  & 0.27 & 0.46 & 0 & 0.22     \\
40$^2$  & 300 & 187 &  4.792 & 4.073 & 29.501 & 40.0 & 0.57 & 10.9 & 2.3 &  0.503 & 9.036 & 62.3  & 0.27 & 0.47 & 0 & 0.20     \\
25  &   0 &  0  &  5.815 & 4.527 & 23.503 & 0    & 0.32 & 0.31 & 0.11 &  0.585 & 13.871 & 0  & 0.52 & 3.2e-4 & 1.8e-2 & 9.9e-3     \\
25  & 300 & 205 &  7.376 & 5.807 & 21.318 & 74.8 & 0.46 & 3.2 & 0.9 &  0.484 & 9.588 & 80.2  & 0.33 & 0.47 & 0 & 0.15     \\
20  & 300 & 227 &  8.602 & 3.105 & 16.927 & 43.7 & 0.38 & 1.6 & 0.5 &  0.792 & 9.245 & 0.04  & 0.49 & 1.3e-3 & 1.4e-2 & 1.3e-2     \\
    &         &         &         &         &     &      &      &      &     &      &      &      &         &    & &        \\
\hline
\multicolumn{17}{ }{ }       \\
\multispan{17}$^1$ Model computed with anisotropic stellar winds and with an $\alpha$--enhanced mixture of the heavy elements.\hfill \\
\multispan{17}$^2$ Model computed with mass loss rates during the WR phase dependent on the metallicity (Crowther et al. 2002).\hfill \\
\end{tabular}
\end{center}

\end{table*}

From Table~\ref{tbl-1} one sees that for $Z=0.040$ the MS lifetimes are 
increased by about 16--27\% when the initial rotational velocity 
increases from 0 to 300 km s$^{-1}$. 
Similar increases
were found at solar metallicity (see papers V and X). 
The He--burning lifetimes are increased by rotation for the 60 and 120 $M_\odot$
models by respectively 12 and 34\%. This comes from the fact that the most massive stars already enter
the WR phase during the core H--burning phase. The strong mass losses that they
experienced at the end of the H--burning phase produce small He--cores at the beginning
of the core He--burning phase. This tends to reduce the central temperatures and to increase
the He--burning lifetimes.

Rotation decreases the He--burning lifetime
for the 25 $M_\odot$ by 8\% at solar metallicity and by 17\% at $Z=0.040$. Contrary to what happens in the high mass star range,  
in the rotating 25 $M_\odot$ stellar model the
He--core at the end of the MS phase is significantly more massive than in the non--rotating model
(by 42\% at $Z=0.040$) and thus the rotating track is much more luminous during most of the He--burning phase
than the non--rotating one. This of course tends to reduce the duration of the core He--burning phase.

\begin{figure}[t]
  \resizebox{\hsize}{!}{\includegraphics{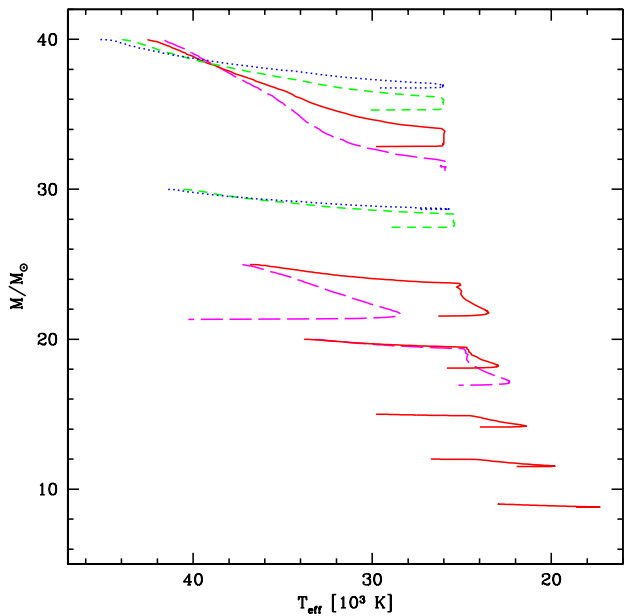}}
  \caption{Evolution of the mass of the stars as a function of the effective temperature for different
initial mass models (from 9 to 40 M$_\odot$) at various metallicities during the Main-Sequence phase
($\upsilon_{\rm ini}=300$ km s$^{-1}$). The initial mass
of the stars is the ordinate of the hottest point of each track (left point). Models
at Z=0.004, 0.008, 0.020 and 0.040 are shown with dotted, short--dashed, continuous and long--dashed lines
respectively. One can note the effect of the crossing of the bistability limit around 25000 K (see text).
}
  \label{bista}
\end{figure}

As indicated in Sect.~2, the present stellar models were computed with the mass loss rates of Vink et al.~(\cite{Vink00}).
This prescription shows a dramatic increase of the mass loss rates for OB Main--Sequence stars when they cross, from
blue to red, the bistability limit which occurs at an effective temperature equal to $\sim$25000~K. 
The increase is due to a drastic change in the ionization of the wind.
To illustrate this effect, the evolution, during
the Main--Sequence phase, of the mass of our stellar models is represented in Fig.~\ref{bista} 
as a function of T$_{\rm eff}$ . When the star models cross the bistability limit, the tracks turn down very abruptly in the cases
of the most massive stars, more smoothly in the cases of smaller initial mass stars.
This results from the important increase of the mass loss rates, by about 1 dex,
when the star crosses the bistability limit. 
As an example, the 25 M$_\odot$ stellar model, which spends only 11\% of its MS lifetime on the red
side of the bistability limit, loses during
this short phase more than 60\% of the mass lost during the whole MS phase.
For stars more massive than about 60 M$_\odot$
and less massive than about 10 M$_\odot$,
there is no jump in the mass loss rates,
since their tracks remain on one side of the bistability limit during the whole Main-Sequence phase. 

One can wonder to what
extent the bistability effect represents an important feature favouring the entry of the stars
in the WR regime. Let us
just remark here that this effect alone is of little help in that respect. Indeed models without rotation
but taking account of this bistability effect give a poor fit of the observed variation
of the number ratio of WR to O--type stars with the metallicity (see Fig.~\ref{wro}), 
while rotating models much better reproduce the
observed data (see Sect.~6).  

\begin{figure}[t]
  \resizebox{\hsize}{!}{\includegraphics{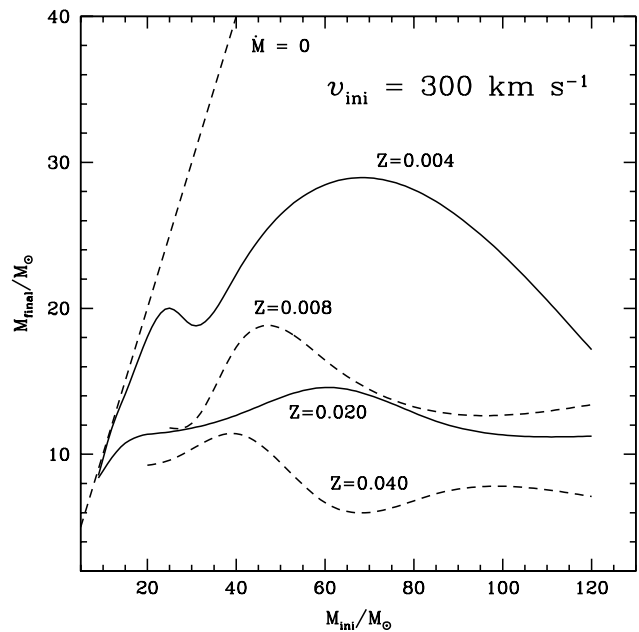}}
  \caption{Relations between the final and the initial mass for rotating stellar models
  at various metallicities. 
The line with slope one, labeled $\dot {\rm M}=0$, corresponds to the case without mass loss.
}
  \label{final}
\end{figure}

From Table~\ref{tbl-1}, comparing the final masses obtained at $Z$ = 0.040 
for the rotating and the non--rotating model, one can see that 
rotation produces in general 
smaller final masses.
Fig.~\ref{final} shows the final masses obtained for different initial masses and metallicities.
At $Z$ = 0.040, most of the stars in the mass range considered here end their life with masses
below 10 $M_\odot$. All stars with masses above about 50 $M_\odot$ reach a final mass between 5 and 7.5 $M_\odot$.
In contrast, at a metallicity one order of magnitude smaller, the final masses of stars are significantly higher,
being in the range between 17 and 29 $M_\odot$ for stars with initial masses above 60 $M_\odot$. This well illustrates
the effect of the metallicity dependence of the mass loss rates, an effect which is qualitatively similar to that found
in non--rotating stellar models (Maeder \cite{Maeder91}).

\begin{figure}[t]
  \resizebox{\hsize}{!}{\includegraphics{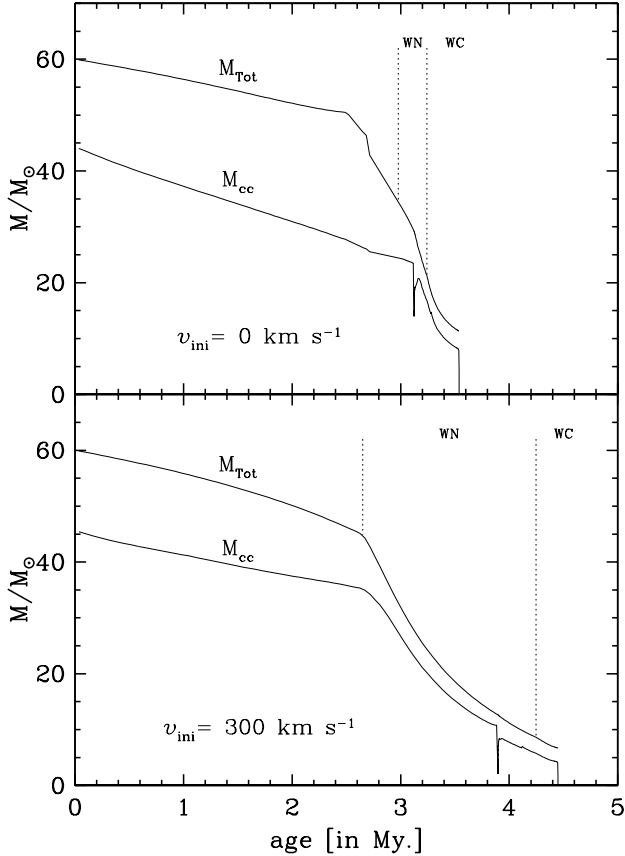}}
  \caption{Evolution as a function of time of the total mass M$_{\rm Tot}$ and of the mass of the convective cores
  M$_{\rm cc}$
  during the H-- and He--burning phases, for a 60  $M_\odot$ stellar model with and
  without rotation at $Z$ = 0.040. The type of the WR star at a given age is given in the upper part of the figure.}
  \label{kipGC}
\end{figure}

\section{Rotating models of Wolf--Rayet stars}

\subsection{Effects of rotation and metallicity on the evolutionary scenarios leading to the formation of WR stars}

The effects of rotation on the evolution of massive single 
stars into the Wolf--Rayet phase  have been discussed by Maeder (\cite{Maeder87}),
Fliegner and Langer (\cite{Fl95}), Maeder \& Meynet \cite{araa} and Meynet (\cite{Me00}).
These studies were mainly based on solar metallicity models. Here we extend 
these discussions to lower and higher than solar metallicities.

We had to choose a set of criteria
to decide when a stellar model enters the WR phase. 
Ideally, of course, the physics of the models should determine 
when the star is a WR star. However our poor knowledge of the physics 
involved, as well as the complexity of models
coupling the stellar interiors to the winds, are such that this 
approach is not yet possible.
Instead, it is necessary to adopt some empirical criteria
for deciding when a star becomes a WR star. 
In this work the star enters the WR phase when two
conditions are fullfilled: 1) 
$\log T_{\rm eff} > 4.0$ (note that after the entry into the WR phase the star
may exceptionally spend very short periods in cooler regions of the HR diagram. In general however,
the whole WR phase is spent at effective temperatures well above $\log T_{\rm eff} = 4.0$);
2) the mass fraction of hydrogen at the surface $X_s$ is inferior to 0.4. 
Reasonable changes to these values (for instance adopting $X_s < 0.3$ 
instead of 0.4) do not affect the results significantly. Let us recall that
the effective temperature considered here is an effective temperature appropriately averaged over the surface
(see Meynet \& Maeder \cite{MMI}). We consider the WR star to be of the eWNL type when the mass fraction
of hydrogen at the surface is superior to 10$^{-5}$, (we adopt here the denomination proposed by Foellmi
et al. \cite{FoL03} for the WNL and WNE phase, based on evolutionary criteria instead of spectroscopic ones,
hence the small ``e'' before the name).
The eWNE phase begins at the end of the eWNL phase, while
the transition WN/WC phase is considered to begin when
the mass fraction of carbon at the surface becomes superior to 10\% of the mass fraction of nitrogen. Both the eWNE
and the transition WN/WC phases end when the WC/WO phase begins, which we assume to begin when
the mass fraction of nitrogen at the surface becomes less than 10\% of the mass fraction of carbon. 
We shall not distinguish here between WC and WO stars. Also, changes of the
numerical values of the limits have very little consequence since the
transitions are fast.

Internal mixing favours the entry into the
WR phase in two ways, firstly
by allowing chemical species produced in the core to diffuse in the radiative envelope and, secondly, by
making the mass of the convective core larger (see also Meynet \& Maeder \cite{MMV}). 
In the non--rotating model,
mass loss by stellar winds is the key physical ingredient which
allows internal chemical products to appear at the surface and thus form a WR star. The star becomes a WR star
only when sufficiently deep layers are uncovered.

In Fig.~\ref{kipGC} the evolution of the structures for 60 M$_\odot$ models with and without rotation
at twice the solar metallicity are shown.  
In the case of the 60 $M_\odot$ model, the most striking differences
between the non--rotating and the rotating model are the following:
\begin{itemize}
\item Inclusion of rotation in the models 
allows an earlier entry into the WR phase.
Typically here in
the non--rotating model the star becomes a WR star when the actual mass is 34.3 $M_\odot$ and 
the mass fraction of H at the centre $X_c$ is 0.04, while in the rotating
model it enters the WR phase when the mass is equal to 44.9 $M_\odot$ and $X_c$ is equal to 0.24. 
\item As a consequence the WR lifetime as well as the duration of the eWNL phase
will be increased by rotation. Also the actual luminosity 
at the entry of the WR phase will be higher for the rotating models.
\item The duration of the WC phase is reduced. 
\end{itemize}
Qualitatively similar effects of rotation were obtained at solar metallicity (cf. paper X) except
for the effect of rotation on the WC phase: at solar metallicity and for the 60 M$_\odot$
stellar model the WC phase is slightly enhanced by rotation (see Table 2 in paper X). At Z = 0.004, the WC phase becomes 
longer when the initial rotation increases (see Table~\ref{tblwr}).
In addition to the above effects, rotation decreases the minimum initial mass of single 
stars going through a WR phase (see Fig.~\ref{wrlifecomp})
and  the duration of the transition WN/WC phase is much longer in the rotating models than in the non--rotating ones.
(see Table~\ref{tblwr}). 

As said in Sect.~2, the importance of these effects depends on the values adopted for the
initial velocity and on various physical ingredients of the stellar models as for instance the mass loss rate and/or the overshooting.
Some illustration of the effects of different values of rotation for 60 M$_\odot$ models
at solar metallicity can be found in Meynet (\cite{Mey99}) and in paper X. Results for a fast rotating 60 M$_\odot$ stellar model
at Z=0.004 ($\upsilon_{\rm ini}=500$ km s$^{-1}$) are presented in Tables~\ref{tbl-1} and \ref{tblwr}. In that last example, we see that
increasing the initial velocity from 300 to 500 km s$^{-1}$ more than doubles the WR lifetime. This enhancement
results from the more efficient internal mixing at high rotation and not to the increase of the mass loss rate due to rotation.

From the present rotating models one can derive two interesting limiting masses. The first,
M$_{\rm OWR}$, is the minimum initial mass of a single star entering the WR phase during the MS phase.
The second, M$_{\rm WR}$, is the minimum initial mass of a single star 
entering the WR phase at any point in the course of its lifetime. These two limiting masses define the mass ranges
of three evolutionary scenarios for the massive stars:
\begin{itemize}
\item For M $>$ M$_{\rm OWR}$, the stars will avoid the Luminous Blue Variable stage
after the MS phase. In this case, they will go through the following
phases: O--eWNL--eWNE--WC/WO. Note that here we assume that
once the star has entered the WR regime, it remains a WR star for the rest of its lifetime. Some stars however
may evolve in cooler regions of the HR diagram after they have entered the WR phase and might thus
encounter the Humphreys--Davidson limit. These star models would present characteristics similar to LBV stars and
could thus belong to this category.
\item For M$_{\rm WR} <$ M $<$ M$_{\rm OWR}$, after the MS phase, the star will evolve
into the cooler part of the HR diagram, where it may encounter the $\Omega\Gamma$--limit
(Maeder \& Meynet \cite{MMVI}) or become a Red Supergiant. In that case, one would have
O--LBV or RSG--eWNL--eWNE--WC/WO. Evolution may not necessarily proceed up to the WC/WO stage,
it may stop at the eWNE or eWNL stage.
\item For M $<$ M$_{\rm WR}$, after the core H--burning phase the O--type star will become 
a supergiant, but it will never enter the WR phase. The blue or red nature of the supergiant
depends among other parameters on rotation (Maeder \& Meynet \cite{MMVII}). The evolutionary
sequence in that case will be O--RSG/BSG. Blue loops may be present especially in the lower
mass star range.
\end{itemize} 
In order to estimate M$_{\rm OWR}$ and M$_{\rm WR}$ we interpolated 
between corresponding evolutionary stages of the tracks
using the logarithm of the initial mass as the interpolating factor.
We obtain the mass limits indicated in Table~\ref{tblml}. 
We can note that, at a given metallicity, the mass limits are lower
when the rotation rate is higher, as expected from the considerations mentioned above. 
As we shall see in Sect.~6, these results on the mass 
limits have interesting consequences for the expected number
of WR stars at different metallicity, as well as for the predicted number
of type Ib/Ic supernovae. 
One notes also from Table~\ref{tblml} that
the mass limits are lower at higher metallicities, as would be the case for models without rotation.
This is due to the
greater mass loss rates experienced by stars at higher metallicity.

\begin{table}[htbp]
\caption{Mass limits for the different evolutionary scenarios (see text).} \label{tblml}
\begin{center}\scriptsize
\begin{tabular}{cccc}
$Z$     &  $\upsilon_{\rm ini}$ &  M$_{\rm WR}$   &   M$_{\rm OWR}$   \\
        & [km s$^{-1}$]       & [$M_\odot$] & [$M_\odot$]   \\ 
        &                       &              &              \\
\hline \\
        &                       &              &              \\
0.040   &  0                    &  29          &  42          \\
0.040   &  300                  &  21          &  39          \\
        &                       &              &              \\
0.020   &  0                    &  37          &  62          \\
0.020   &  300                  &  22          &  45          \\
        &                       &              &              \\
0.008   &  300                  &  25          &  69          \\
        &                       &              &              \\
0.004   &  300                  &  32          &  75          \\
        &                       &              &              \\
\hline                              
\end{tabular}
\end{center}

\end{table}

The mass range for which the
WR phase is preceded by a LBV and/or a RSG phase shifts to higher
values at lower metallicity, it also extends over a larger mass interval.
Typically, at $Z$ = 0.040 rotating models predict that 
such a scenario for single massive stars
occurs in the mass range between 21 and 39 $M_\odot$, while it occurs
in the mass range between 32 and 75 $M_\odot$ at $Z$ = 0.004.
Since qualitatively a same trend is predicted by the non--rotating models,
this result would imply that the upper luminosity of LBV stars
should be lower at higher metallicity.
However one should be cautious here. First, one cannot
discard the possibility that a star enters the WR phase and later undergoes
a shell ejection similar to a LBV outburst due for instance to reaching
the $\Omega\Gamma$--limit (see Maeder \& Meynet~\cite{MMVI}). Moreover the observed
stellar populations are a mixture of stars of different initial velocities, whose initial
distributions may depend on the metallicity.
This would contribute to the blurring of the schematic picture just described above.

As already stated in Sect.~2, the results 
shown is Table~\ref{tblml} are sensitive to various
physical ingredients. For instance,
an enhancement of the core sizes and/or of the initial velocity and/or of the mass loss rates
would favour the WR formation and thus
the lowering of these mass limits. Thus the numbers shown in Table~\ref{tblml}
may undergo some changes in the future when improvements in these physical ingredients will be reached.
However, at least qualitatively, the trend they show, namely the fact that rotation lowers
the mass limits, will likely persist. This is well illustrated by the fact that a similar trend was already
present in our paper V models which were computed with different prescriptions for the mass loss rates, the
overshooting and the rotational mixing (see Sect.~2). 

For a given initial mass and an initial rotational velocity, the greater the metallicity the larger
the luminosity range spanned during the WR phase, and a lower minimum luminosity is reached. This can be seen
from Fig.~\ref{dhr004}.
Interestingly we can see that at a given metallicity, the higher initial mass stars are in general the progenitors
of the less luminous WC stars. This is mainly due to the fact that the mass loss rates increase with the luminosity,
enabling the most massive stars to enter the WR phase at an earlier stage and thus to lose mass at high
rates during a larger portion of their lifetime. Thus these stars end with small final masses and hence with small
luminosities since at this stage they are following a well known mass luminosity relation
(Schaerer \& Maeder \cite{SC92}).


\subsection{The Wolf--Rayet lifetimes}

\begin{figure}[t]
  \resizebox{\hsize}{!}{\includegraphics[angle=-90]{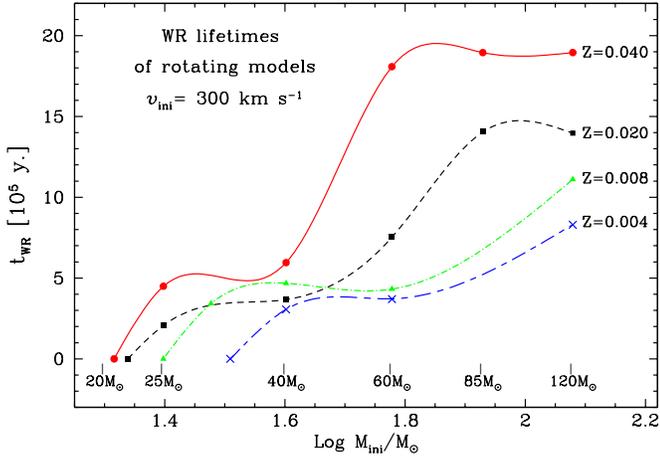}}
  \caption{Lifetimes of Wolf--Rayet stars from various initial masses at four different metallicities.
All the models begin their evolution with $\upsilon_{\rm ini}$ = 300 km s$^{-1}$ on the ZAMS, the 
corresponding time--averaged velocities during the MS O--type star phase are given in Table~\ref{tbl-1}.}
  \label{wrlife}
\end{figure}

The WR lifetimes of the present rotating models for the four metallicities 
are plotted as a function of the initial mass in Fig.~\ref{wrlife}. As was the case with the non--rotating models (Maeder~\cite{Maeder91}; Maeder \& Meynet \cite{MM94}), 
the metallicity
dependence of the mass loss rates is responsible for two features: 1) for a given initial mass and velocity 
the WR lifetimes are greater at higher metallicities. Typically at Z=0.040
and for M $>$ 60 $M_\odot$ the WR lifetime is of the order of 2 Myr, while at the metallicity of the
SMC the WR lifetimes in this mass range are between 0.4--0.8 Myr; 2) the minimum mass 
for a single star to evolve into the WR phase is lower at higher metallicity. 

\begin{figure}[t]
  \resizebox{\hsize}{!}{\includegraphics[angle=-90]{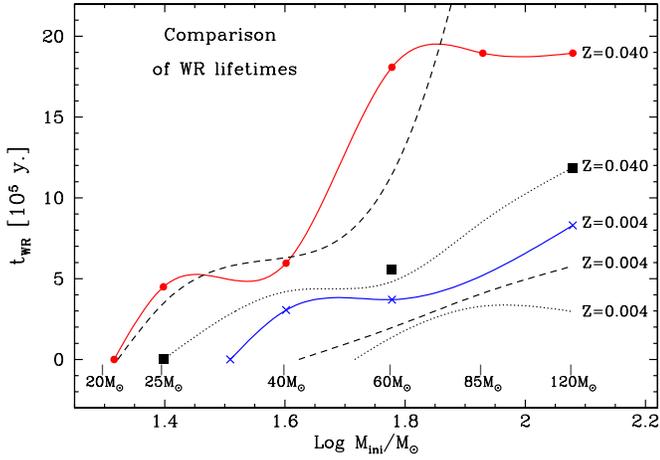}}
  \caption{Lifetimes of Wolf--Rayet stars from various initial masses for $Z$ = 0.040 and 0.004
  from different sets of models: continuous lines are for the present rotating models
  with $\upsilon_{\rm ini}$ = 300 km s$^{-1}$, the dotted  and dashed lines are for the non--rotating stellar models
  with ``normal'' and ``enhanced'' mass loss rates computed by Meynet et al. (\cite{Mey94}). The black squares
  indicate the lifetimes obtained for the non--rotating 25, 60 and 120 $M_\odot$ models obtained in the present work
  at $Z=0.040$.}
  \label{wrlifecomp}
\end{figure}

It is interesting to compare the WR lifetimes obtained at $Z$ = 0.040 and 0.004 from different sets of models (see
Fig.~\ref{wrlifecomp}). First, one can notice that the present non--rotating models with the updated mass loss rates
give very similar results to those obtained with the ``normal'' mass loss rates by Meynet et al. (\cite{Mey94}; see
how the black squares give a good fit to the $Z$ = 0.040 dotted curve). 
As was discussed in paper X, rotation has a similar effect as an enhancement of the mass loss rates
on the WR lifetimes. Namely, for a given initial mass it increases the WR lifetime and also lowers 
the minimum initial mass of single stars going through a WR phase. Note however that quantitatively 
the results obtained from rotating models may be different from the one obtained from the enhanced 
mass loss rate models. At low metallicity, the curve
obtained from the rotating models is always above the one obtained from the enhanced mass loss rate models.
More importantly, rotation seems to be much more efficient in lowering the value of M$_{\rm WR}$.
This is due to the fact that rotation helps in forming WR stars through internal mixing, a process which,
in contrast to mass loss rates, is important for all massive stars, not only the most massive ones.
At high metallicity, the situation is somewhat different, because
when metallicity increases the mass loss rate becomes the dominant parameter.

\begin{figure*} [t]
\resizebox{\hsize}{!}{\includegraphics[angle=-90]{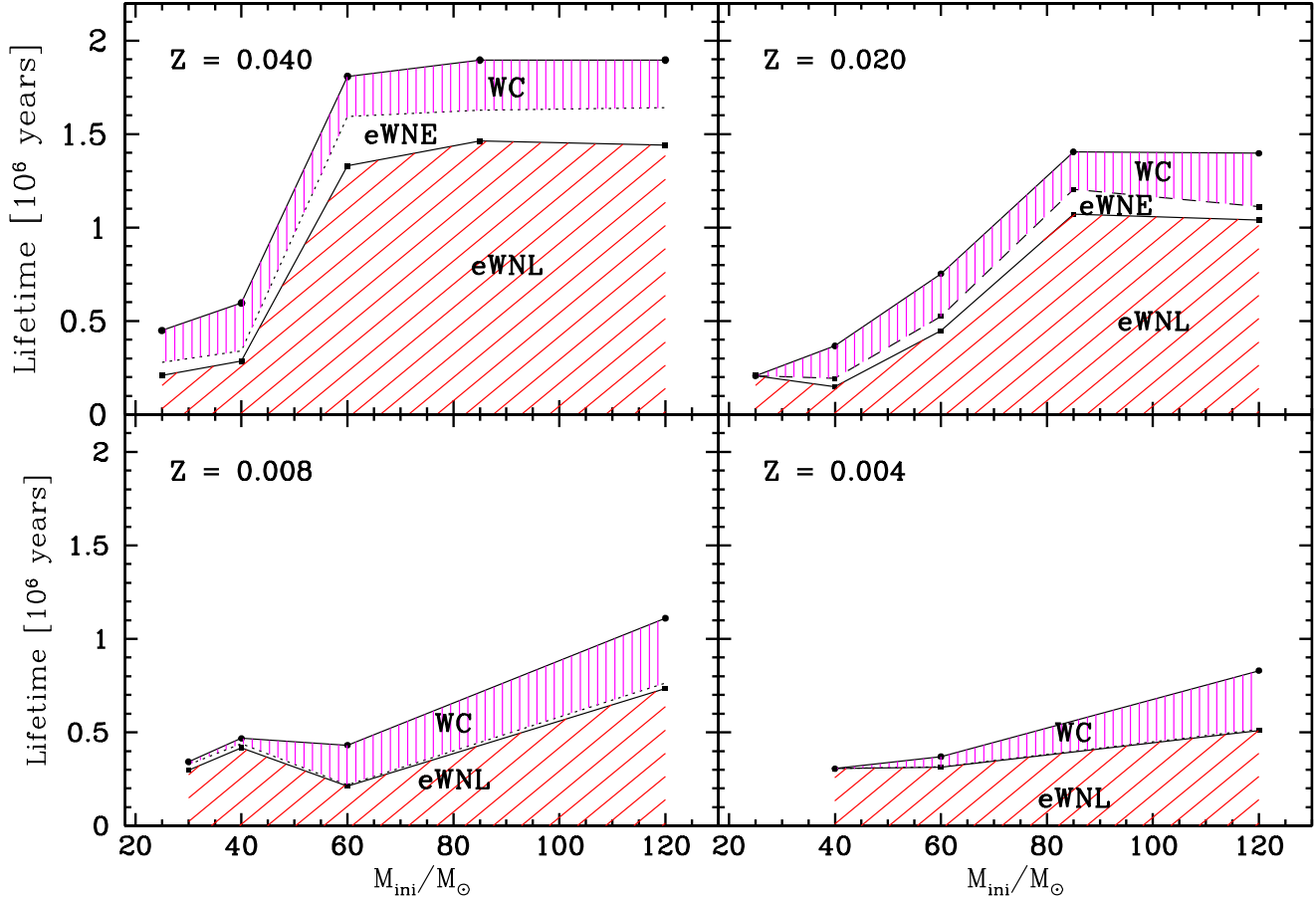}}
\caption{Variation of the durations of the WR subphases as a function of the initial mass
at various metallicities. All the models begin their evolution with $\upsilon_{\rm ini}$ = 300 km s$^{-1}$.}
  \label{tma}
\end{figure*}

In Fig.~\ref{tma}, the duration of the different WR subphases is plotted as a function of the initial mass for
the various 
metallicities considered in this work. Only the results with rotation are plotted.
The durations of these different subphases are given in Table~\ref{tblwr}.
The greatest part of the WR lifetime is spent in the eWNL phase. Note that, as explained above, rotation increases the duration of this phase. 
For identical initial rotational velocities, the duration of the eWNL phase is greater at higher metallicity.
This results from the metallicity dependence of the mass loss rates. At higher metallicity, 
the higher mass loss rates by stellar winds enable the star to enter the WR phase at an earlier
stage. The eWNE phase is also longer at higher metallicity
(as is also the case for non--rotating models).
The WC phase keeps more or less
the same duration for all the metallicities in the higher mass star range. In the lower mass star
range the WC phase is longer at higher metallicity as a result of the shift towards a lower value of the minimum
initial mass of single stars needed to become a Wolf--Rayet star.

As already explained in paper X,  
in the rotating stellar models a new phase of modest,
but non--negligible duration, appears: the so--called transition WN/WC phase
(see the values of $t_{\rm WN/WC}$ given in Table~\ref{tblwr}). 
This phase is characterized by the simultaneous presence at the surface of both
H-- and He--burning products. The reason for this
is the shallower chemical gradients which build up inside the rotating models.
These shallower gradients inside the stars also produce a smoother evolution of the surface abundances 
as a function of time (see Fig.~\ref{cs40}). 
As explained in paper X,
for a transition WN/WC phase to occur, it is necessary 
to have---for a sufficiently long period---both 
an He--burning core and a CNO--enriched envelope.
In general, in the highest mass stars, mass loss removes the CNO--enriched envelope 
too rapidly
to allow a long transition WN/WC phase to occur. 
In the low mass range, the time spent in the WR phase is 
too short and the H--rich envelope too
extended to allow He--burning products to diffuse up to the surface. 
Consequently, a significant transition WN/WC phase only appears in an intermediate mass range 
between $\sim$ 30 and 60 $M_\odot$ for $v_{\rm ini}=300$ km s$^{-1}$.
In general the duration of this phase also increases significantly when mass loss rates depending on the metallicity are used during
the WR phase (see Sect~7).  

\begin{table}[t]
\caption{Lifetimes in Myr of WR stars of different initial masses and metallicities.
The durations of various WR subphases are also indicated as well as 
theoretical predictions for various number ratios.} \label{tblwr}
\begin{center}\scriptsize
\begin{tabular}{ccccccc}
M   & $v_{\rm ini}$        & $t_{\rm WR}$ & $t_{\rm eWNL}$ & $t_{\rm eWNE}$ & $t_{\rm WN/WC}$ & $t_{\rm WC}$  \\
    &                   &          &           &           &             &           \\
\hline
    &                   &          &           &           &             &           \\
\multispan{7}\hfill {\bf Z=0.004} \hfill \\
    &                   &          &           &           &             &           \\
120 & 300               & 0.8298   & 0.5105    & 0.0066    & 0.0089      & 0.3127    \\
 60 & 300               & 0.3704   & 0.3126    & 0.0039    & 0.0113      & 0.0539    \\ 
 60 & 500               & 0.8617   & 0.4617    & 0.1422    & 0.1089      & 0.2578    \\ 
 40 & 300           & 0.3057   & 0.3057    & 0.0000    & 0.0000      & 0.0000    \\
    &                   &          &           &           &             &           \\
\multispan{7}\hfill {\bf Z=0.008} \hfill \\
    &                   &          &           &           &             &           \\
120 & 300               & 1.1107   & 0.7354    & 0.0282    & 0.0167      & 0.3471    \\
 60 & 300               & 0.4315   & 0.2113    & 0.0088    & 0.0186      & 0.2114    \\ 
 40 & 300               & 0.4677   & 0.4168    & 0.0238    & 0.0372      & 0.0271    \\    
 30 & 300          & 0.3421   & 0.2971    & 0.0299    & 0.0528      & 0.0151    \\    
    &                   &          &           &           &             &           \\
\multispan{7}\hfill {\bf Z=0.040} \hfill \\
    &                   &          &           &           &             &           \\
120 &     0             & 1.1829   & 0.7646    & 0.1276    & 0.0034      & 0.2907    \\
120 & 300               & 1.8956   & 1.4398    & 0.2004    & 0.0048      & 0.2554    \\
120$^1$ & 300           & 2.3190   & 1.7200    & 0.4339    & 0.0382      & 0.1651    \\
 85 &     300           & 1.8955   & 1.4626    & 0.1641    & 0.0980      & 0.2688    \\
 85$^1$ & 300           & 2.2710   & 1.6190    & 0.3087    & 0.1723      & 0.3433    \\   
 60     & 0             & 0.5592   & 0.2083    & 0.0552    & 0.0009      & 0.2957    \\ 
 60 &     300           & 1.8083   & 1.3294    & 0.2640    & 0.0083      & 0.2149    \\ 
 60$^1$ & 300           & 2.1382   & 1.5450    & 0.4515    & 0.0940      & 0.1417    \\ 
   40 &   300           & 0.5960   & 0.2857    & 0.0537    & 0.0291      & 0.2566    \\    
 40$^1$ & 300           & 0.5710   & 0.2221    & 0.0433    & 0.0231      & 0.3056    \\    
 25     & 300           & 0.4493   & 0.2091    & 0.0706    & 0.0602      & 0.1696    \\ 
\hline
    &                   &          &           &           &             &           \\
 \multispan{7}\hfill Predicted number ratios \hfill \\
    &                   &          &           &           &             &           \\
 $v_{\rm ini}$        & ${\rm WR} \over {\rm O}$       & ${\rm eWNL}\over {\rm WR}$   
 & ${\rm eWNE}\over {\rm WR}$   & ${\rm WN/WC}\over {\rm WR}$  & ${\rm WC}\over {\rm WR}$  & ${\rm WC}\over{\rm WN}$$^2$     \\
                      &          &           &           &             &        &           \\
\hline
    &                   &          &           &           &             &           \\
\multispan{7}\hfill {\bf Z=0.004} \hfill \\
                      &           &           &            &             &         &            \\
    300               & 0.02      & 0.83      & 0.00       & 0.01        & 0.15    &  0.18      \\
                      &           &           &            &             &         &            \\
\multispan{7}\hfill {\bf Z=0.008} \hfill \\
                      &           &           &            &             &         &            \\
    300               & 0.05      & 0.74      & 0.05       & 0.07        & 0.22    &  0.28      \\
                      &           &           &            &             &         &            \\
                      &           &           &            &             &         &            \\
\multispan{7}\hfill {\bf Z=0.020$^3$} \hfill \\
                      &           &           &            &             &         &            \\
      0               & 0.02      & 0.35      & 0.16       & 0.00        & 0.49    &  0.97      \\ 	      
    300               & 0.07      & 0.66      & 0.05       & 0.04        & 0.25    &  0.35     \\
                      &           &           &            &             &         &            \\		      
\multispan{7}\hfill {\bf Z=0.040} \hfill \\
                      &           &            &            &              &         &           \\
    0                 & 0.06      & 0.46       & 0.10       & 0.00         & 0.44    &  0.78      \\ 
    300               & 0.16      & 0.63       & 0.12       & 0.05         & 0.24    &  0.32     \\
    300$^1$           & 0.18      & 0.59       & 0.15       & 0.06         & 0.26    &  0.36     \\
                      &           &            &            &              &         &           \\		      		      	      
\hline   
                     &          &           &           &             &        &           \\                       
\multispan{7}$^1$ Model computed with mass loss rates during the WR phase dependent \hfill \\
\multispan{7}on the metallicity (Crowther et al. \cite{Cro02}).\hfill \\
\multispan{7}  \\
\multispan{7}$^2$ ${\rm WC}\over{\rm eWNL+eWNE}$.\hfill \\
\multispan{7}  \\
\multispan{7}$^3$ Values obtained in paper X.\hfill \\
\end{tabular}
\end{center}

\end{table}

\begin{figure}[t]
  \resizebox{\hsize}{!}{\includegraphics[angle=-90]{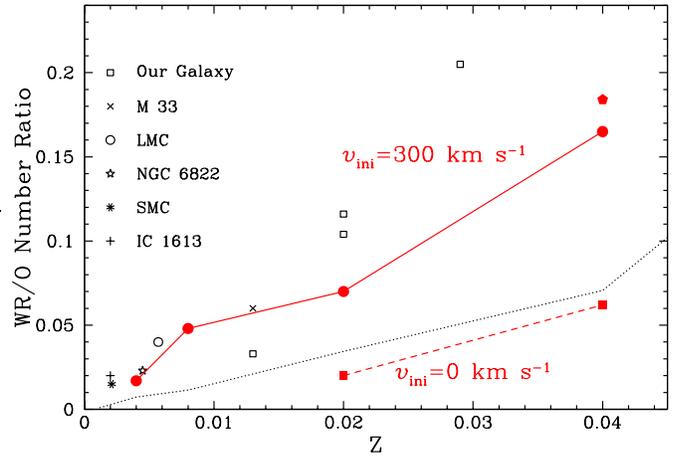}}
  \caption{Variation of the number ratios of Wolf--Rayet stars to O--type stars as a function of the metallicity.
The observed points are taken from Maeder \& Meynet (\cite{MM94}).
The dotted line shows the predictions of the models
of Meynet et al. (\cite{Mey94}) with normal  mass loss rates.
The continuous and the dashed lines show the predictions of the present rotating and 
non--rotating stellar models
respectively. The black pentagon shows the ratio predicted by Z=0.040 models computed
with the metallicity dependence of the mass loss rates during the WR phase proposed
by Crowther et al.~(\cite{Cro02}).
}
  \label{wro}
\end{figure}

\begin{figure}[t]
  \resizebox{\hsize}{!}{\includegraphics[angle=-90]{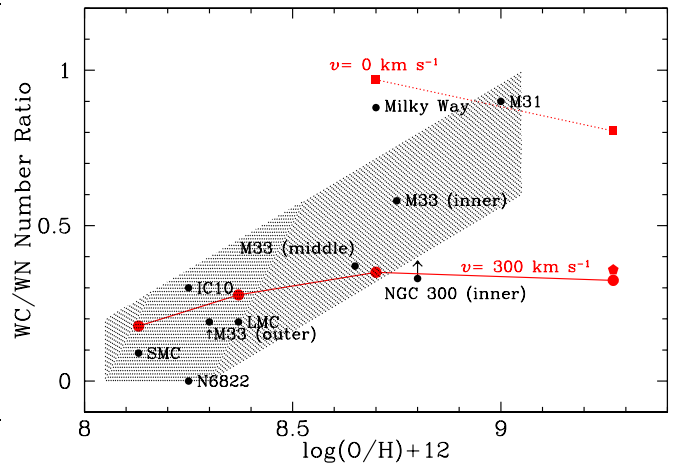}}
  \caption{Variation of the number ratios of WN to WC stars as a function of metallicity.
The black circles are
observed points taken  from Massey \& Johnson (\cite{Mas98} and see references therein), except
for the SMC (Massey \& Duffy \cite{Mas01}), for NGC 300 (Schild et al. \cite{Sch02})
and for IC10, for which we show the estimate from Massey \& Holmes (\cite{Mas02}) .
The continuous and dotted lines show the predictions of the present rotating and non--rotating stellar models
respectively. The black pentagon shows the ratio predicted by Z=0.040 models computed
with the metallicity dependence of the mass loss rates during the WR phase proposed
by Crowther et al.~(\cite{Cro02}).}
  \label{massey}
\end{figure}

\begin{figure}[t]
  \resizebox{\hsize}{!}{\includegraphics[angle=-90]{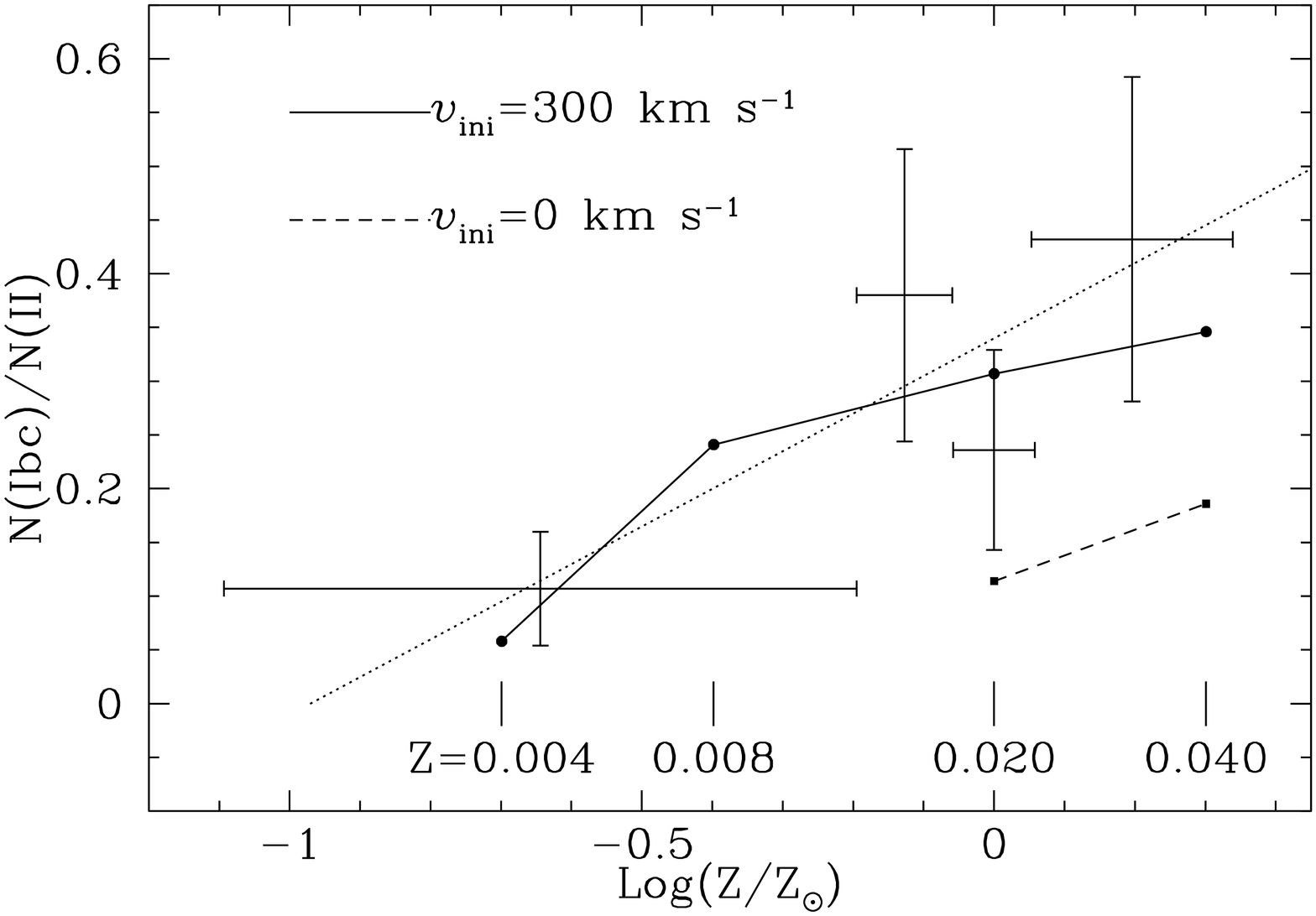}}
  \caption{Variation of the number ratios of type Ib/Ic supernovae to type II supernovae. The crosses
  with the error bars correspond to the values deduced from observations by Prantzos \& Boissier (\cite{Pr03}). The
  dotted line is an analytical fit proposed by these authors. The continuous and dashed line show the predictions of 
  the present rotating and non--rotating stellar models. }
  \label{ibcii}
\end{figure}

\section{Comparison with the observed properties of Wolf--Rayet star populations}

Following Maeder (\cite{Maeder91}) and
Maeder \& Meynet (\cite{MM94}), one can easily estimate the theoretical
number ratio of WR to O--type stars in a region of constant star formation.
This ratio is simply given by the ratio of the averaged lifetimes of a WR star to that
of an OV--type star. The averaged lifetimes are weighted means of the lifetimes over 
the initial mass function (IMF).
Assuming a Salpeter IMF slope (${\rm d}N/{\rm d} M\propto M^{-(1+x)}$, with $x$ = 1.35), considering the O--type and WR star lifetimes 
given in Tables~\ref{tbl-1} and \ref{tblwr} and the mass limits $M_{\rm WR}$ given in Table~\ref{tblml}
\footnote{At $Z$ = 0.004, the value for the minimum initial mass of O--type stars
is taken from the rotating models of Maeder \& Meynet (\cite{MMVII}). At $Z$ = 0.008
we use the value obtained from the ``enhanced'' mass loss rate models
of Meynet et al. (\cite{Mey94}; see Table 2 in Maeder \& Meynet \cite{MM94}).},
we obtain the predicted ratios given at the bottom
of Table~\ref{tblwr}.

Let us emphasize here that in the present work
we assume that the $\upsilon_{\rm ini}$ = 300 km s$^{-1}$
stellar models are well representative of the behaviour of the majority of the OB stars.
In future studies, when information on the 
distribution of the initial rotation velocities
will be  available at different metallicities, it will certainly be interesting 
to convolve the theoretical
results obtained for various initial velocities with these observed
velocity distributions.

Looking at the results presented in Table~\ref{tblwr}, one can note that rotating models predict that
\begin{itemize}
\item The fraction of WR stars with respect to O--type stars increases with the metallicity.
\item The fraction of eWNE with respect to WR stars increases with the metallicity, as does the fraction of WC stars
(although the ratio seems to saturate around 0.25 for $Z \ge 0.020$). 
\item The fraction of eWNL decreases with the metallicity again reaching a saturation level
around 0.65 for $Z \ge 0.020$.
\item The fraction of stars in the transition WN/WC phase has a non--monotonic behaviour, increasing
from 0.01 at $Z$ = 0.004 to 0.07 at $Z$ = 0.008, and decreasing to values between 0.04--0.05 for
$Z \ge$ 0.020.
\end{itemize}

Comparisons with observed number ratios are shown in Figs.~\ref{wro} and \ref{massey}.
The values for the non--rotating models (both those computed with the present physics and those computed by
Maeder \& Meynet \cite{MM94}) are well below the observed values. 
The ratios predicted by the models with rotation are
in much better agreement with
the observations.
In particular the ratios predicted by the rotating models  are in quite good agreement 
with the observed ones at low metallicity. They give predictions that are well in line with the results obtained by Foellmi et al. (\cite{FoS03}, \cite{FoL03}).
Let us recall that these authors
concluded that, in contrast to previous expectations, the binary channel for WR star formation does not seem to be a favoured scenario at low metallicity. 
Indeed, as already mentioned in the introduction, the fraction of binaries
among the WR stars of the Small (40\%) and Large (30\%)  Magellanic Cloud appears to be similar to that found in our Galaxy.
Moreover, not all WR stars in a binary system must necessarily owe their existence to a Roche Lobe Overflow, 
since the two stars in a binary system may be sufficiently distant from each other to prevent that such a process occurs.
Thus models from single massive stars should be able to reproduce more than half the number of observed WR stars in the Magellanic Clouds, a
condition quite well fulfilled by the rotating models and not realized by the non--rotating ones.
At high metallicity, the agreement is less good, although
still much better than the one obtained from non--rotating models. 
Let us note that the samples
at high metallicity do not have the same
level of completeness as for the Magellanic Clouds.
In the Clouds, the distance and metallicity are relatively well known quantities, moreover
the internal extinction is relatively weak and thus does not constitute a severe barrier. 
The central regions of galaxies and of the Milky Way in particular do not offer such
good conditions, and completeness problems are likely much more
important than in the Magellanic Clouds. 

In Fig.~\ref{massey}, the predicted values for the WC/WN number ratio are compared
with observations in regions supposed to have undergone a constant star formation rate. 
The observations show that
the WC/WN ratio increases with the metallicity
along a relatively well defined relation.
The observed point
for the solar neighborhood is however well above the general trend. According to Massey 
(\cite{Mas03}) this may result from an underestimate of the number of WN stars.

At low metallicity, models with rotation are well within the general observed trend. At solar
metallicity they are just at the inferior limit and at twice the solar metallicity they
are below the extrapolated observed trend. The non--rotating models predict
(at least for $Z \ge$ 0.020) significantly higher WC/WN ratios. If the value predicted
for the solar metallicity lies well above the real observed value at this metallicity
(see Massey \cite{Mas03}), the one at twice the solar metallicity appears to be in good
agreement with the extrapolated observed trend.

Taken at face value, it seems that rotating models overestimate
the number of WN stars at high metallicity. 
Apart from invoking completeness problems in the observed sample, how might the comparisons between the ratios
predicted by the rotating models and the observed ones be improved ? We see two possible solutions:
1) From Fig.~\ref{massey}, it clearly appears that non--rotating stars, or more reasonably,
stars rotating with relatively small velocities at high metallicity, would make the predicted points
remain in the general observed trend. Slow rotators undergo little rotational mixing, thus they
keep steep gradients of chemical composition in their interior. The fraction of the mass of the star
presenting abundances characteristic of those of WN stars is therefore smaller than
in a rotating star. This tends to decrease the WN lifetime. 
2) The use of higher mass loss rates during the WN phase will tend to remove the H--rich envelope more rapidly,
decreasing the WN lifetime and favouring an early entry into the WC phase.

The first solution is not reasonable for at least two reasons. Firstly, Fig.~\ref{wro} shows that
slowly rotating models are unable to account for the high fraction of WR stars observed at high metallicity.
Secondly the observed number ratio of type Ib/Ic supernovae with respect to type II supernovae would also be severely
underestimated by non--rotating models (Prantzos \& Boissier \cite{Pr03}, see also Sect. 6.1 below).
The second solution, enhanced mass loss rates during the WN phase, appears more promising, although the effects
of such an enhanced mass loss rate may be different according to the stage at which the star
enters the WR phase.
The most massive stars already enter the WN phase during the MS phase. An enhancement of the mass loss rates
in that case can have the opposite effect of the one desired, namely it may lengthen the WN phase. 
Indeed a higher mass loss produces a more important reduction of the convective H--burning core, 
accompanied by a severe reduction of the central temperature. This tends to
make the end of the H--burning phase longer, and
thus to eventually increase the duration of the WN phase. Only an enhancement of the mass loss rates for those WN stars beyond the H--burning phase would
reduce the WN phase. 

Recently Crowther et al.~(\cite{Cro02}) proposed a metallicity dependence for the mass loss rates of WR stars
comparable to that predicted for O--type stars, i.e. with $(Z/Z_\odot)^{1/2}$. 
We have recomputed the $Z$ = 0.040 models between 40 and 120 $M_\odot$ with this new prescription. 
The lifetimes of the different WR subphases are given in Table~\ref{tblwr}.
We note that for the 120 $M_\odot$ star, which enters the WR phase during the core H--burning phase, the stronger
mass loss rates given by the new prescription lead to an increase of the WN phase
and a decrease of the WC phase. The reason is just the one invoked above. In the case of the 40 $M_\odot$ model, where
only 17\% of the WN phase is spent in the core H--burning phase, the metallicity dependent mass loss rates during
the WR phase give a shorter WN phase and a slightly longer WC phase. It is likely that in the case of the
25 $M_\odot$ model, which spends its whole WR phase in the post core H--burning phase, the reduction of the WN phase
and the enlargement of the WC phase would be more important. Supposing that for this 25 $M_\odot$ stellar model the total WR lifetime is not
much changed by the use of metallicity dependent mass loss rates during the WR phase, one can estimate the durations
of the WN and WC phases one would obtain by simply assuming that the new $t_{\rm WN}$ is equal to that
given in Table~\ref{tblwr} divided by $(0.04/0.02)^{1/2}=1.4$ and the new $t_{\rm WC}$ is equal to $t_{\rm WR}-t_{\rm WN}$. 
If one does this one finds that the new $t_{\rm WN}$ and $t_{\rm WC}$ are 0.20 and 0.25 Myr
respectively. With these values, one obtains the predicted number ratios given in the last line of Table~\ref{tblwr}.
One sees that the use of metallicity
dependent mass loss rates during the WR phase does not much increase the predicted value for the
WC/WN ratio at high metallicity (see also Fig.~\ref{massey}). 

Thus in view of the above discussion, the two possibilities which presently appear as the most probable for
reconciling the predicted WC/WN ratio with the observed ones are 1) that the observed WC/WN ratio is overestimated
as seems to be the case at least for the Milky Way (Massey~\cite{Mas03}); 2) that the mass loss rates during the 
WN phase after the core H--burning phase are higher than presently assumed. One may also wonder
what is the role played by binaries in this context.

\subsection{The type Ib/Ic supernovae}

Current wisdom associates the supernovae of type Ib/Ic with the explosion of WR stars, the
H--rich envelope of which has been completely removed either by stellar winds and/or by mass transfer through Roche Lobe overflow in a close binary system. 
If we concentrate on the case of single star models, 
theory predicts that the fraction of type Ib/Ic supernovae
with respect to type II supernovae should be higher at higher metallicity. The reason is the same
as the one invoked to explain the increasing number ratio of WR to O--type stars with the metallicity $Z$, namely the
growth of the mass loss rates with $Z$. Until recently very little observational evidence has been found 
confirming this predicted behaviour. The situation began to change with the work by Prantzos \& Boissier~(\cite{Pr03}). These authors
have derived from published data the
observed number ratios of type Ib/Ic supernovae to
type II supernovae for different metallicities. The regions considered  are regions of constant star formation rate. 
Their results are plotted in Fig.~\ref{ibcii}. Using the 
predicted ratios at solar metallicity deduced from our rotating models (paper X), they concluded that
rotating models were much better able than non--rotating ones to reproduce the observed ratio at solar
metallicity. 

We complement here the discussion of Prantzos \& Boissier~(\cite{Pr03}) by comparing theory to the observation at other metallicities.
Theoretical 
ratios are simply given by the ratio $N_{\rm Ib/Ic}/N_{II}$, where $N_{\rm Ib/Ic}$ is equal to
the integration of the IMF over the range 
of M$_{\rm WNE}$ to 120 $M_\odot$, where M$_{\rm WNE}$ is the minimum initial mass of the stars that end their lives as eWNE or WC stars,
and where $N_{II}$ is the integration of the IMF between 8 and M$_{\rm WNE}$.
To determine the values of M$_{\rm WNE}$ we used the same technique as the one used to determine
M$_{\rm WR}$ and M$_{\rm OWR}$. We obtained 20.6, 22, 25 and 52 for $Z$ = 0.040, 0.020, 0.008, 0.004 respectively.
Looking at Fig.~\ref{ibcii}, it clearly appears that the conclusion of  Prantzos \& Boissier~(\cite{Pr03})  
is confirmed by the comparisons at low and high metallicities. Rotating models
give a much better fit to the observed data than non--rotating models. This comparison can be viewed as a check
of the lower initial mass limit M$_{\rm WNE}$ of the stars evolving into a WR phase without hydrogen, 
while the comparison between the observed and
predicted number ratio of WR to O--type stars involves not only the value of the minimum 
initial mass of stars evolving into the WR phase but also the durations of the WR phase. 
In that respect the comparison with the supernovae ratios is a more direct check of the correctness of the value of $M_{\rm WNE}$
which at high metallicity appears rather close to M$_{\rm WR}$.  

\section{Evolution of the surface abundances}

\begin{figure}[tbp]
  \resizebox{\hsize}{!}{\includegraphics{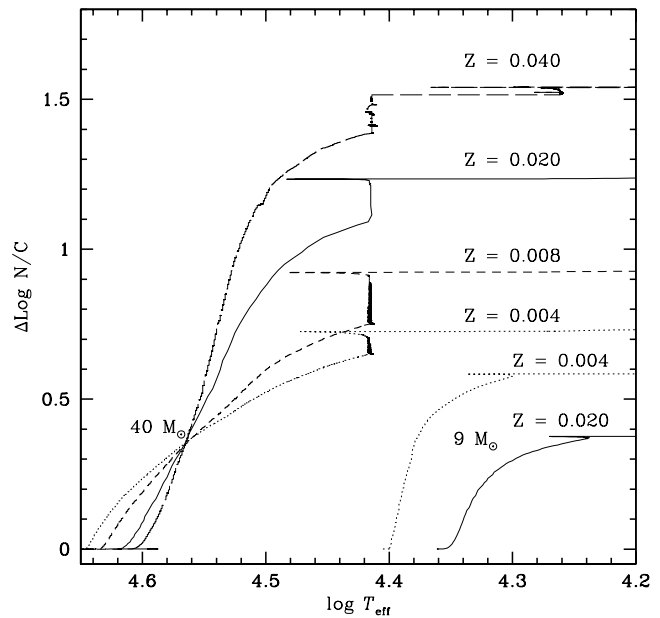}}
  \caption{Evolution during the MS phase of the N/C ratios (in number) at the surface of 
  rotating stellar models as a function of the effective temperature. 
  The differences in N/C ratios are given with respect to the initial values. }
  \label{nc}
\end{figure}

\begin{figure*}[tbp]
  \resizebox{\hsize}{!}{\includegraphics{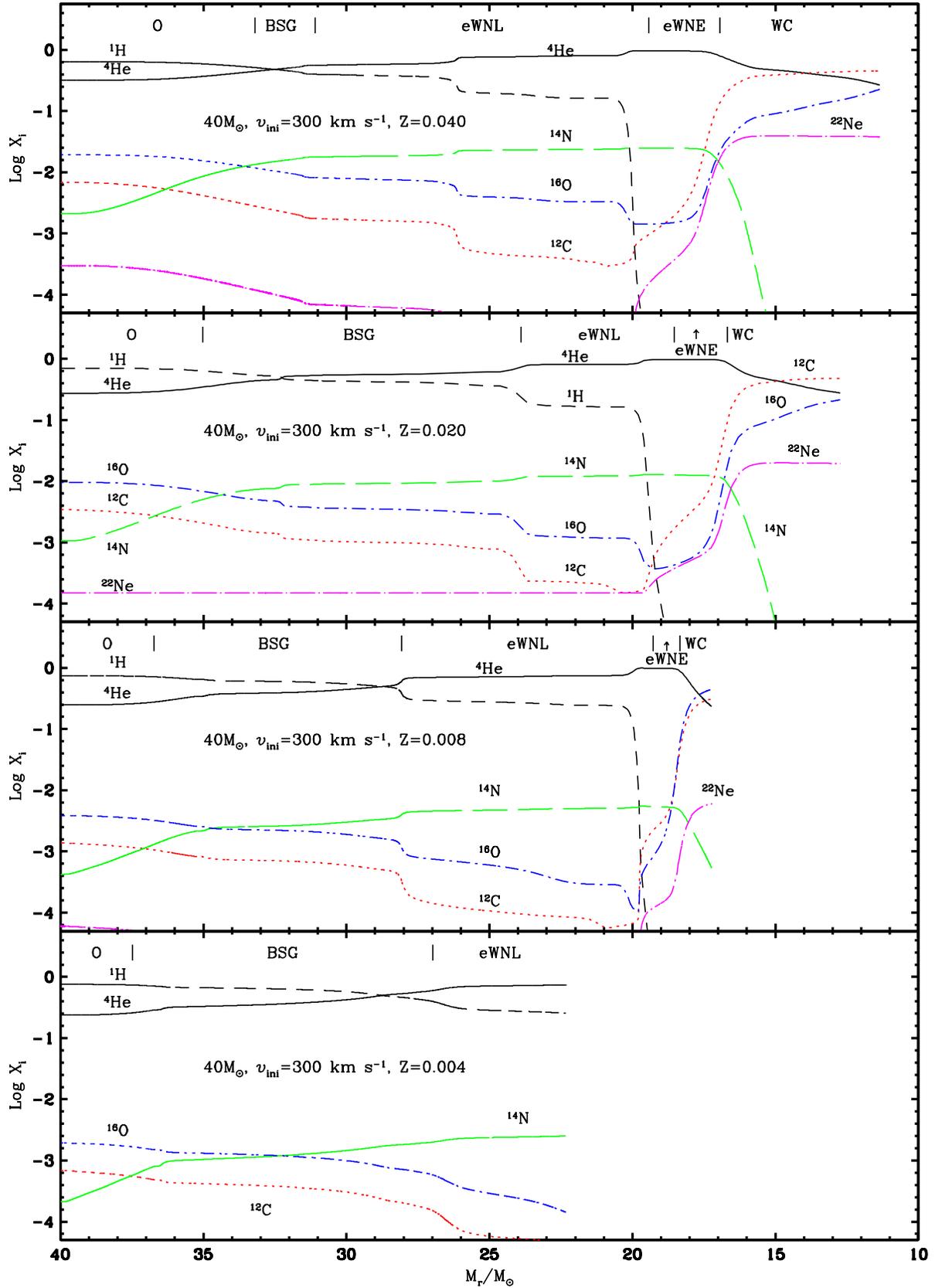}}
  \caption{Evolution as a function of the actual mass of the abundances (in mass fraction) 
  of different elements at the surface of  rotating  40 $M_\odot$ stellar models
  at various metallicities.
}
  \label{cs40}
\end{figure*}

The effects of
rotation on the surface abundance of massive stars at solar composition have been 
discussed by 
Heger \& Langer (\cite{he00}), Meynet \& Maeder (\cite{MMV}; \cite{MMX}). One of the main features
is the enhancement 
of the N/C ratios at the surface during the MS phase.
In papers VII and VIII 
we obtained that, for a given initial mass and
velocity, the surface enrichment induced by rotation is higher at lower metallicity.
This is a consequence of the fact that at lower metallicity stars are more compact and 
have steeper internal gradients of $\Omega$ (see paper VII and VIII). 
This implies strong shear mixing of the chemical elements and explains why in Fig.~\ref{nc} the N/C ratio
at the surface of the 9 $M_\odot$ model at $Z=0.004$ is higher than for the corresponding model at $Z=0.020$.

For the more massive stars the situation is different. Indeed one can see in Fig.~\ref{nc} that the N/C ratio
at the end of the MS phase at the surface of the rotating 40 $M_\odot$ at Z=0.040 is higher than the N/C ratio
obtained at the same stage at the surface of the 40 $M_\odot$ at Z=0.004. 
Why is the behaviour of the most massive stars so different from that of the less massive ones ?
The reason is that, in the high mass star range, stellar winds become the dominant effect.
High mass losses facilitate the surface
enrichment  by uncovering deep layers whose chemical composition has been changed
by nuclear processing and/or rotational mixing. 

Fig.~\ref{cs40} shows the evolution of the surface abundances in rotating models with an initial mass of 40 M$_{\odot}$. 
Before commenting on the effects of the metallicity, let us briefly recall the effects of rotation. In rotating
models one can note the following differences with respect to the non--rotating ones:
\begin{itemize}
\item In rotating models 
the progressive changes
of the abundances of CNO elements from the initial cosmic values  
to the values of the nuclear equilibrium of the
CNO cycle are smoother. This is due to rotational mixing which smoothes internal
chemical gradients.
\item Due to rotational mixing, the change
of abundances also occurs  much earlier in the peeling--off process.
This is also true for the changes of 
H and He. We note that
the nuclear equilibrium CNO values are essentially model independent as already stressed
a long time ago (Smith \& Maeder \cite{SmithM91}). This is true whether H is still 
 present or not. 
\item  In the rotating case, the transitions between the eWNE phase and the WC stage are smoother,
so that there are some stars observed in the  transition state which 
correspond to the so--called  WN/WC stars, (Conti \& Massey \cite{CM89}; Crowther et al. \cite{Cr95}).
These stars show simultaneously some $^{12}$C and $^{14}$N. They could also
have some  $^{22}$Ne excess. Since the attribution of spectral types
is a complex matter, it may even be that some of the stars in the 
transition stage are given a spectral type WNE or WC, thus we might well
have a situation where a WN star would have some  $^{22}$Ne excess
or a WC star would still have some $^{14}$N present.
\item At the entry into the WC phase, the $^{12}$C and  $^{16}$O abundances are lower in rotating models, 
and the abundance of He is higher.
\item Also, the fraction of the WC phase spent with lower C/He and O/He ratios is longer 
in models with rotation.
\item However, rotation does not affect the high level of the $^{22}$Ne abundance
during the WC phase. This is a consequence of the fact that most of 
this $^{22}$Ne results from the transformation of the $^{14}$N produced 
by the CNO cycle in the previous H--burning core. The value of the $^{14}$N 
and therefore that of the $^{22}$Ne is fixed by the
characteristics of the CNO at equilibrium, which in turn depends on 
the nuclear physics and not on the pecularities of the stellar models. 
It is interesting
to mention here that the high overabundance of $^{22}$Ne at 
the surface of the WC star predicted by the models 
is well confirmed by the observations (Willis \cite{Wi99}; Dessart et al. \cite{De00}).
\end{itemize}

Looking at Fig.~\ref{cs40}, one can see how the evolution of the surface abundances is affected
by rotation at four different metallicities.
Among the most striking effects one can note:
\begin{itemize}
\item The eWNE phase is more extended at high metallicity than at low metallicity (see also Fig.~\ref{tma}).
This is the result of the higher mass loss rates experienced at high metallicity.
The removal of the outer layers by stellar winds is more rapid, therefore when the star enters the eWNE phase
it does so at an earlier stage of the core He--burning phase, when the core is less massive. Thus the He--rich envelope
extends over a larger fraction of the star.
\item For the same physical reason as above, the star enters the WC phase at an earlier stage
of the core He--burning phase  at high metallicity.
Therefore at high metallicity one expects that the surface abundances will be characterized by
higher He abundance and larger C/O ratios than at low metallicity. 
This has interesting consequences for the WC populations expected at various
$Z$ (see below).
\item The lower the metallicity, the less mass is lost by stellar winds
and thus the less evolved is the stage at which the evolution ends.
The 40 $M_\odot$  stellar model at Z=0.004 never enters the WC phase. 
This model will explode as a core collapse type II supernova.
\end{itemize}

\begin{figure}[tbp]
  \resizebox{\hsize}{!}{\includegraphics{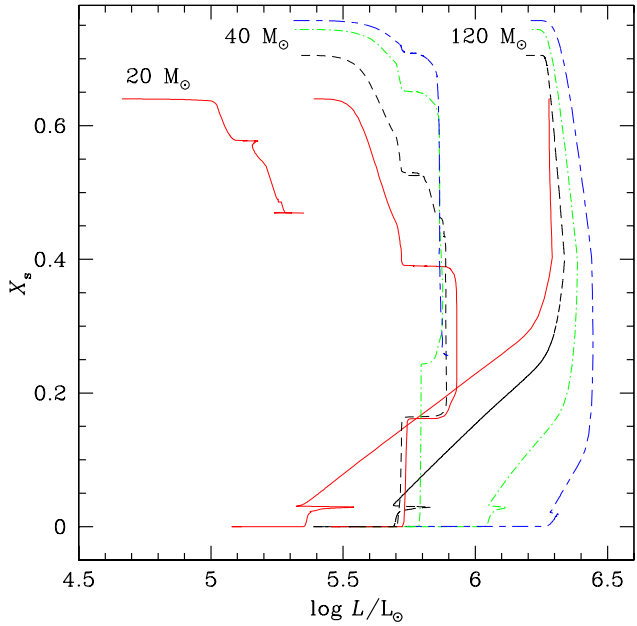}}
  \caption{Evolutionary tracks in the X$_{\rm s}$ versus log {\it L}/L$_\odot$ plane,
where X$_{\rm s}$ is the hydrogen mass fraction at the surface. The initial masses are indicated.
Long--short dashed curves show the evolution of $Z$ = 0.004 models, dashed--dotted curves, short--dashed curves
and continuous lines show the evolutions for $Z$ = 0.008, 0.020 and 0.040 respectively.}
  \label{xsl}
\end{figure}
 
\begin{figure}[tbp]
  \resizebox{\hsize}{!}{\includegraphics{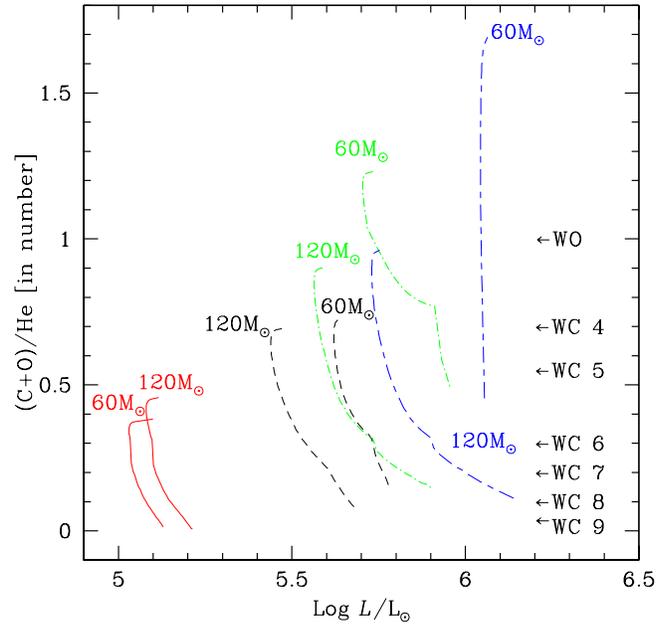}}
  \caption{Evolution of the ratios (C+O)/He as a function of the luminosity at the surface of 60 and 120 $M_\odot$ rotating models for various initial 
  metallicities (see text). Long--short dashed curves show the evolution of $Z$ = 0.004 models, dashed--dotted curves, short--dashed curves
and continuous lines show the evolutions for $Z$ = 0.008, 0.020 and 0.040 respectively.
The correspondence between the (C+O)/He ratios and the different WC subtypes as given by Smith \& Maeder (\cite{SmithM91}) is indicated on the right of the figure.
}
  \label{cohe}
\end{figure}

Fig.~\ref{xsl} presents 
the evolution of the surface H--content as a function of the luminosity.
The tracks go downwards in this diagram. Firstly there is an initial 
brightening without surface H--depletion, then due to the concomitant effects of mixing and mass
loss, the surface abundance of hydrogen decreases. The most striking effect of the metallicity
is seen for the 120 $M_\odot$ stellar models. At higher metallicity, one notes that
the high mass loss prevents the track to go through the zone characterized by high luminosity
and small surface hydrogen abundance. Only the models at low metallicity explore this domain
of the $X_s$ versus Log $L/L_\odot$ plane.

The evolution of the (C+O)/He ratio as a function of luminosity is shown
in Fig.~\ref{cohe}, which is the key diagram for WC stars as shown by 
Smith \& Maeder (\cite{SmithM91}). The track for a given mass goes up
to the left. Indeed as a function of time, surface abundances characteristic of a more advanced
He--burning stage appear at the surface resulting in an increase of the (C+O)/He ratio. At the same time mass loss reduces the actual mass of the star and therefore
its luminosity.
Since in rotating models there is a very progressive change of the surface abundance, all the models
would seem to enter the WC phase with more or less the same (C+O)/He ratio. But this would be misleading
since this transition phase is very short. Thus we decided here to plot the evolution of
the surface ratios corresponding to the last 90\% of the WC lifetime, so that the figure can give a better idea
of the types of WC stars formed at the different metallicities. 

One sees that at high metallicity, mainly late type WC stars are present, while at low metallicity, the late type
WC stars are absent and earlier types are present. 
This is quite consistent with the observational
fact that late type WC stars are only found in high metallicity regions (Smith \& Maeder \cite{SmithM91}).
Fig.~\ref{cohe} also shows that
a given WC subtype is reached at lower luminosity for a higher metallicity, a fact which is also
consistent with the observed trend (Crowther et al. \cite{Cro02}). Such a behaviour results from 
the smaller
mass at the entry in the  WC phase for models at higher metallicity. The smaller
mass in its turn is due mainly to the much longer WN phase experienced by the high metallicity stars, during which
intense stellar winds peel off the star.

\section{Conclusion}

Rotation gives much better results than non--rotating models for the
following observed features: 
\begin{itemize}
\item the observed number ratio of WR to O--type stars for metallicities between 0.004 and 0.040, 
\item the observed ratio of WN to WC stars 
for metallicities lower than solar, 
\item the observed fraction of WR stars in the transition WN/WC phase 
\item and the observed ratio of type Ib/Ic 
to type II supernovae at different metallicities. 
\end{itemize}
Only in the case of the WC/WN ratio observed at high metallicity there may be
a difficulty, although completeness problems may be part of the problem. 

Interestingly,   
the features which were already well reproduced by non--rotating models, such 
as the  abundance of $^{22}$Ne at the surface of WC stars or
the fact that late type WC stars are only observed at high metallicity,
are also well accounted for
by the present rotating models. 
This work further confirms that stellar rotation is an essential ingredient of
massive star evolution, both for the MS phase and for the advanced stages
like the WR stages and the Supernovae. We may therefore also anticipate that
the nature of the remnants will be different depending on rotation and metallicity.
In forthcoming papers we shall use the present rotating models to estimate the contribution
of the WR stars to the synthesis of $^{26}$Al in the Galaxy, and also to address
the question of the rotation rates of pulsars and the progenitors of the collapsars
suggested by Woosley (\cite{Wo93}) to be possible progenitors of long gamma ray bursts.

\end{document}